\documentclass[%
reprint,
nofootinbib,
nobibnotes,
amsmath,amssymb,
aps,
prl,
]{revtex4-2}

\usepackage{bm}
\usepackage{mathtools}
\usepackage[dvipsnames]{xcolor}
\usepackage{graphicx}
\usepackage{enumitem}
\usepackage[bookmarksnumbered=true,bookmarksopen=true]{hyperref}

\hypersetup{
  colorlinks,
  linkcolor=NavyBlue,
  citecolor=PineGreen,
  urlcolor=PineGreen
}

\definecolor{SoftRose}{HTML}{DC9CA1}
\definecolor{HardBlue}{HTML}{4A8FD4}

\newcommand{\dd}{\mathrm{d}}
\newcommand{\ii}{\mathrm{i}}
\newmuskip\pFqmuskip
\newcommand{\pFqcomma}{\mskip\pFqmuskip}
\newcommand*\pregFq[6][8]{
  \begingroup
  \pFqmuskip=#1mu\relax
  \mathcode`\,=\string"8000
  \begingroup\lccode`\~=`\,
  \lowercase{\endgroup\let~}\pFqcomma
  {}_{#2}\mathbf{F}_{#3}{\left[\genfrac..{0pt}{}{#4}{#5};#6\right]}
  \endgroup
}
\newcommand*\pFq[6][8]{
  \begingroup
  \pFqmuskip=#1mu\relax
  \mathcode`\,=\string"8000
  \begingroup\lccode`\~=`\,
  \lowercase{\endgroup\let~}\pFqcomma
  {}_{#2}F_{#3}{\left[\genfrac..{0pt}{}{#4}{#5};#6\right]}
  \endgroup
}
\newcommand*\MeijerG[8][8]{
  \begingroup
  \pFqmuskip=#1mu\relax
  \mathrm{G}^{#2,#3}_{#4,#5}\!
  \mathcode`\,=\string"8000
  \begingroup\lccode`\~=`\,
  \lowercase{\endgroup\let~}\pFqcomma
  \left[\genfrac..{0pt}{}{#6}{#7};#8\right]
  \endgroup
}

\begin{document}

\hbadness=10000

\title{Silent Heavy Fields: An Exclusion Rule for the Cosmological Collider}

\author{Shuntaro Aoki\textsuperscript{1}}
\email{shuntaro.aoki@riken.jp}
\author{Zhehan Qin\textsuperscript{2}}
\email{zq242@cam.ac.uk}
\author{Masahide Yamaguchi\textsuperscript{3,4,5}}
\email{gucci@ibs.re.kr}
\author{Yuhang Zhu\textsuperscript{3}}
\email{yhzhu@ibs.re.kr}

\affiliation{\vspace{6pt}\textsuperscript{1}RIKEN Center for Interdisciplinary Theoretical and Mathematical Sciences (iTHEMS), Wako, Saitama 351-0198, Japan}
\affiliation{\vspace{3pt}\textsuperscript{2}Department of Applied Mathematics and Theoretical Physics, University of Cambridge, Wilberforce Road, Cambridge, CB3 0WA, UK}
\affiliation{\vspace{3pt}\textsuperscript{3}Cosmology, Gravity and Astroparticle Physics Group, Center for Theoretical Physics of the Universe, Institute for Basic Science, Daejeon 34126, Korea}
\affiliation{\vspace{3pt}\textsuperscript{4}Department of Physics, Institute of Science Tokyo, 2-12-1 Ookayama, Meguro-ku, Tokyo 152-8551, Japan}
\affiliation{\vspace{3pt}\textsuperscript{5}Department of Physics and IPAP, Yonsei University, 50 Yonsei-ro, Seodaemun-gu, Seoul 03722, Korea}

\date{\today}

\begin{abstract}
  The cosmological collider uses oscillatory signatures in primordial correlators to probe heavy fields during inflation. This probe has blind spots, as heavy fields can keep ticking in time while remaining silent in the correlators. In this \textit{letter}, we derive an exclusion rule under which cosmological collider signals vanish. The rule covers two classes of diagrams where a de Sitter invariant heavy sector couples covariantly to a single external field at one vertex. We first prove it for exchanges mediated by the two-point function of the heavy sector to all loop orders, using a spectral decomposition followed by a field redefinition. The result then extends to one-loop polygons by matching their oscillatory contributions onto those of the first class. The approximate signals of previously studied processes that fall into these classes must therefore be revisited. As a nonvanishing example, we derive the leading signal of a triangle loop for the first time in a toy model motivated by chemical potentials.
\end{abstract}
\maketitle

\section{Introduction}
Heavy fields present during inflation oscillate with frequencies set by their masses. Through interactions with the light perturbations, their evolution from one interaction event to another can leave distinctive imprints on primordial correlators, taking the form of oscillations in the logarithm of momentum ratios. Such imprints provide the physical basis for cosmological collider (CC) physics \cite{Chen:2009zp,Baumann:2011nk,Noumi:2012vr,Arkani-Hamed:2015bza}, giving us access to the heavy particle spectrum at the inflationary scale.

Over the past decade, extensive work on tree-level cosmological correlators has led to substantial theoretical progress \cite{Arkani-Hamed:2018kmz,Baumann:2019oyu,Sleight:2019mgd,Sleight:2019hfp,Sleight:2020obc,Pajer:2020wxk,Jazayeri:2021fvk,Goodhew:2020hob,Melville:2021lst,Pimentel:2022fsc,Jazayeri:2022kjy,Wang:2022eop,Qin:2022fbv,Qin:2023ejc,De:2023xue,Arkani-Hamed:2023bsv,Stefanyszyn:2023qov,Aoki:2024uyi,Liu:2024str,Qin:2025xct,Werth:2024mjg,Chen:2024glu,Stefanyszyn:2024msm,Pimentel:2025rds,Cheung:2025dmc,Chowdhury:2025ohm,Jazayeri:2025vlv,Cheung:2026wrz,Huenupi:2026abj,Arundine:2026myr,Ferreira:2026awc,Belrhali:2026ktb,Belrhali:2026ygh,Grafe:2026avi,Arkani-Hamed:2017fdk}, with the resulting predictions guiding most current observational searches for CC signals \cite{Cabass:2024wob,Goldstein:2024bky,Sohn:2024xzd,Suman:2025vuf,Suman:2025tpv,Philcox:2025wts,Anbajagane:2025uro,Green:2026yev,Kumar:2026ogn,Kumar:2026dih,Philcox:2026tjj}. More recently, one-loop processes have attracted increasing attention \cite{Xianyu:2022jwk,Salcedo:2022aal,Lee:2023jby,Qin:2023bjk,Cespedes:2023aal,AguiSalcedo:2023nds,Qin:2024gtr,Bhowmick:2025mxh,Jain:2025maa,Cespedes:2025ple,Baumann:2025qjx,Pimentel:2026kqc,Grafe:2026qsm,Farren:2026hao,Westerdijk:2025ywh,Baumann:2026atn,Chen:2026dqp,Aoki:2026vbc,Liu:2026jzn,Qin:2026yuo,Borinsky:2026roj}, as they offer richer phenomenology~\cite{Chen:2016uwp,Chen:2016hrz,Chua:2018dqh,Li:2019ves,Lu:2019tjj,Hook:2019vcn,Wang:2019gbi,Li:2020xwr,Wang:2020ioa,Aoki:2020zbj,Cui:2021iie,Ema:2023dxm,Bodas:2025vpb,You:2026xoq} and can even provide the leading contribution to CC signals in some scenarios \cite{Chen:2018xck,Hook:2019zxa,Tong:2023krn,Aoki:2026olh,You:2026xoq}. This is particularly relevant when the particles involved must be produced in pairs, as is the case for fermions \cite{Chen:2016uwp,Chen:2016hrz,Hook:2019vcn,Hook:2019zxa,Chen:2018xck,Cui:2021iie,You:2024hit,Han:2024qbw,You:2026xoq,Aoki:2026vbc} and particles carrying conserved charges \cite{Wang:2019gbi,Sou:2021juh}.

Existing studies of loop phenomenology often evaluate the CC signal approximately through a late-time expansion of the heavy propagators. However, the oscillating modes that appear in this expansion do \textit{not} by themselves guarantee a nonzero CC signal. A particularly sharp illustration of this was recently found in a fermion bubble with a Yukawa coupling, which stays completely silent in the bispectrum once the fermion propagation respects de Sitter (dS) symmetry \cite{Aoki:2026vbc}. An analogous null was also found for a nonderivative scalar bubble in \cite{Aoki:2026gge}. These examples raise a broader question: under what conditions do heavy fields that tick in time remain silent in the correlators?

In this \textit{letter}, we derive an exclusion rule that answers this question for two broad classes of diagrams, in which a dS invariant heavy sector couples locally and covariantly to a single external field at one vertex. In the first class, the heavy sector connects this vertex to the rest of the diagram through its bulk two-point function. Using a spectral decomposition \cite{Marolf:2010zp,DiPietro:2021sjt,Sleight:2021plv,Loparco:2023rug}, we express this two-point function as a superposition of tree exchanges, with the covariant coupling at this vertex mapped onto a quadratic mixing between the external field and the exchanged heavy mode. This mixing can be removed by a field redefinition \cite{Cespedes:2025ple,Aoki:2026vbc}, so that the associated CC signal vanishes. The result holds irrespective of the interactions at the other end of the two-point function, which need not respect dS symmetry. The simplest realisation of this class is an elementary tree exchange, but the argument extends nonperturbatively to fully interacting composite two-point functions of the heavy sector.

In the second class, the heavy fields form a one-loop $N$-gon, for which the absence of CC signals is not manifest. In the soft limit of the external field, the candidate nonanalytic contribution comes from the region in which the two internal propagators attached to its vertex carry soft loop momentum, while the remaining subgraph stays hard. Order by order in the soft expansion, we match this hard subgraph onto local operators at a single vertex \cite{Qin:2026yuo}, reducing the oscillatory contributions of the $N$-gon to those of bubble diagrams of the first class. The exclusion rule therefore applies, and the CC signals naively expected from the oscillations of the internal heavy fields at late times are absent from the final correlator.

The exclusion rule has immediate phenomenological implications. Previously studied loop processes \cite{Li:2020xwr,Chen:2016hrz,Ema:2023dxm,Lu:2019tjj,Cui:2021iie,Aoki:2020zbj} that fall into these two classes must have vanishing CC signals under the assumptions of the rule, in contrast to earlier approximate estimates. A nonzero signal therefore requires an ingredient that evades these assumptions, such as a noncovariant coupling to the external field or a heavy sector that breaks dS symmetry. Motivated by the helicity dependent Whittaker propagation induced by \emph{helical chemical potentials} during inflation \cite{Wang:2019gbi,Sou:2021juh,Qin:2022fbv}, we illustrate the latter possibility with a heavy scalar that shares this Whittaker propagation, while keeping its coupling to the external field covariant. Restoring realistic spin would mainly introduce the corresponding helicity and tensor structures without changing the underlying mechanism. We present the first analytical calculation of the leading CC signal from a Whittaker triangle, and find a signal that is generically nonzero and vanishes continuously as dS covariance is restored. 

\vspace{2pt}
\noindent\textbf{Notations and conventions.} We work in the Poincar\'e patch of four dimensional dS space with metric $g_{\mu\nu}=a^2(\tau)\eta_{\mu\nu}$ and $a(\tau)=-1/(H\tau)$ for $\tau<0$. The external field $\phi$ can actually be any light field coupled perturbatively to the heavy sector in our discussion. Observationally it corresponds to the massless inflaton fluctuation, although in explicit calculations we take it to be conformally coupled, as this simplifies the expressions without affecting our main discussion. Bold symbols denote comoving momentum with $k=|\bm k|$. We consider the generic single soft limit $k_L/k_S\to0$, taking the magnitude $k_L$ of one external momentum to be small compared with a fixed hard scale $k_S$ while the hard momentum approaches a fixed configuration consistently with momentum conservation, so that no other partial sum of external momentum becomes soft. By a CC signal we mean any term in this expansion that oscillates in $\log(k_L/k_S)$ with a nonzero frequency.

\begin{figure}[t]
  \centering
  \includegraphics[width=\columnwidth]{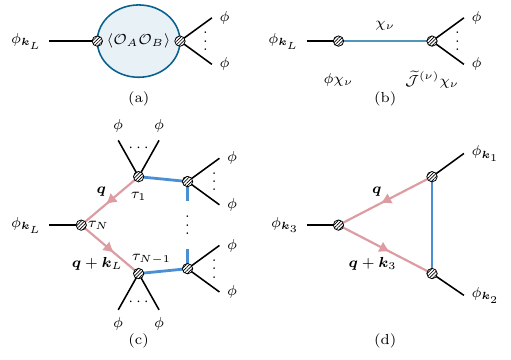}
  \caption{ Diagrams covered by the exclusion rule. (a) A bulk two-point exchange diagram with a single soft external leg $\phi_{\bm k_L}$. The blob represents an arbitrary connected dS covariant two-point function of the heavy sector. (b) One scalar principal series component of (a), represented by $\chi_\nu$ exchange. (c) A general one-loop polygon with $N$ vertices. (d) The triangle contribution to the three-point correlator, with $\bm k_3$ soft. In (c) and (d), \textcolor{SoftRose}{red} and \textcolor{HardBlue}{blue} lines denote soft and hard heavy propagators, respectively. The soft lines carry $\bm q$ and $\bm q+\bm k_L$, with $\bm k_L=\bm k_3$ in (d). In (c), $\tau_N$ labels the soft vertex, while $\tau_1$ and $\tau_{N-1}$ label the hard endpoints of the propagators carrying $\bm q$ and $\bm q+\bm k_L$, respectively. The region relevant to the CC signal has $|\tau_1|,|\tau_{N-1}|\ll|\tau_N|$. }
  \label{fig:selection-rule}
\end{figure}

\section{Vanishing CC signals}
We now turn our attention to the central question of this \textit{letter}, namely under what conditions heavy fields that tick in time remain silent in the correlators. Its answer is summarised at the outset in the following rule.

\vspace{5pt}
\noindent\textbf{Exclusion rule.} \textit{For diagrams with a single external field $\phi_{\bm k_L}$ at one vertex, the CC signal generated by the heavy sector vanishes in the following two classes,}
\begin{enumerate}[label=(\roman*), leftmargin=1.5em]
\item \textit{diagrams with this vertex connected to the other vertex through a bulk two-point function of the heavy sector at arbitrary loop order, as in Fig.~\ref{fig:selection-rule}(a), in any kinematic configuration,}
\item \textit{one-loop polygons, as in Fig.~\ref{fig:selection-rule}(c), at every order in the generic single soft expansion,}
\end{enumerate}
\noindent\textit{provided that the heavy sector has dS covariant intrinsic dynamics and couples to this external field through a local dS covariant interaction, assuming the Bunch-Davies vacuum. Arbitrary local couplings are allowed at other vertices.}\footnote{The rule applies separately to each permutation of the external legs, provided that the corresponding vertex satisfies the stated conditions.}
\vspace{5pt}

In what follows, we refer to the vertex carrying $\phi_{\bm k_L}$ as the soft vertex and to the remaining vertices as hard vertices. Let us first prove the rule for bulk two-point exchange, and then show how one-loop polygons reduce to this case by matching their oscillatory terms onto bubble diagrams.

\subsection{Bulk two-point exchange}

Consider the bulk two-point diagram in Fig.~\ref{fig:selection-rule}(a). The local dS covariant interaction at the soft vertex contracts derivatives of $\phi$ with tensor operators $\mathcal O_A^{\mu_1\cdots\mu_r}$, built from the heavy fields and their covariant derivatives. We can always integrate by parts to move all derivatives from $\phi$ onto these operators,\footnote{Integration by parts modifies the action only by boundary terms. These contribute at most local/contact terms and therefore do not affect the spatially nonlocal CC signal considered here.} so that each term takes the form
\begin{align}
  S_{A}&= g_A\int \dd^4x\sqrt{-g}\,\mathcal O_A^{\mu_1\cdots\mu_r} \nabla_{\mu_1}\cdots\nabla_{\mu_r}\phi\nonumber\\
  &= g_A\int \dd^4x\sqrt{-g}\,\phi\,\mathcal O_A\,,
\end{align}
where $\mathcal O_A\equiv(-1)^r\nabla_{\mu_r}\cdots\nabla_{\mu_1} \mathcal O_A^{\mu_1\cdots\mu_r}$ is a local covariant scalar operator. The heavy fields may be bosons or fermions of arbitrary spin. In all cases, the soft vertex reduces to a coupling between $\phi$ and a local covariant scalar operator of the heavy sector.

The remaining external legs may couple through an otherwise arbitrary local interaction, which we write as
\begin{align}
  S_B=g_B\int \dd^4x\sqrt{-g}\,\mathcal J_B[\{\phi\};\tau]\cdot\mathcal O_B\,.
\end{align}
Here $\mathcal O_B$ is a local covariant operator of the heavy sector, while $\mathcal J_B$ describes its otherwise arbitrary local coupling to the remaining external fields. The notation $\{\phi\}$ includes these fields and their time and spatial derivatives, with any additional scale factors absorbed into the arbitrary time-dependent coefficients of $\mathcal J_B$. The dot denotes contraction over the heavy tensor components. At order $g_Ag_B$, the heavy dynamics enter only through the connected two-point function $\langle\mathcal O_A(X)\mathcal O_B(Y)\rangle_c$. We keep this correlator fully nonperturbative, so the argument applies at arbitrary loop order within the heavy sector.

Since $\mathcal O_A$ is a scalar operator, the dS spectral decomposition \cite{Marolf:2010zp,DiPietro:2021sjt,Sleight:2021plv,Loparco:2023rug} gives the principal series contribution to this correlator in the form
\begin{align}
  &\left\langle\mathcal O_A(X)\mathcal O_{B,\mu_1\cdots\mu_\ell}(Y)\right\rangle_c\nonumber\\
  &= \int_0^\infty \dd\nu\,\rho_{AB}(\nu)\,\mathcal D_{\mu_1\cdots\mu_\ell}^{(\nu)}(Y)\,\mathcal G^{(\nu)}(X,Y)+\cdots\,.\label{spectral}
\end{align}
Here $\mathcal G^{(\nu)}$ is the propagator of a scalar field $\chi_\nu$ with mass $m_\nu$, where $m_\nu^2/H^2=9/4+\nu^2$. The covariant differential operator $\mathcal D_{\mu_1\cdots\mu_\ell}^{(\nu)}$ acts at $Y$ and supplies the tensor structure of the heavy operator $\mathcal O_B$. The ellipsis denotes other contributions that do not carry the oscillatory signal of interest\footnote{The full correlator may contain UV and late-time IR divergences, but these do not affect our statement about the CC signal. UV divergences are removed by local counterterms, while late-time IR growth associated with local contact terms can be separated from the finite contribution from massive exchange, as illustrated in \cite{Cespedes:2025ple}.}. Equation~\eqref{spectral} holds for each Schwinger-Keldysh (SK) component, with the indices suppressed. Integrating by parts transfers the derivatives in $\mathcal D^{(\nu)}$ to $\mathcal J_B$, leaving a local coupling to $\chi_\nu$ with coefficient
\begin{align}
  \widetilde{\mathcal J}^{(\nu)}[\{\phi\};\tau]\equiv\left(\mathcal D^{(\nu)}\right)^\dagger\cdot\mathcal J_B[\{\phi\};\tau]\,,
\end{align}
where $(\mathcal D^{(\nu)})^\dagger$ is the adjoint differential operator obtained by integration by parts. Each spectral component therefore reduces to the well-understood tree-level scalar exchange shown in Fig.~\ref{fig:selection-rule}(b), with interactions
\begin{align}
  S_{\text{tree}}=\int\dd^4x\sqrt{-g}\,\left(g_A\phi\chi_\nu+g_B\widetilde{\mathcal J}^{(\nu)}[\{\phi\};\tau]\chi_\nu\right)\,.
\end{align}
This linear mixing at the soft vertex generates no CC signal, irrespective of the hard coupling. This is most easily seen by making the local field redefinition
\begin{align}
  \widetilde\chi_\nu=\chi_\nu-\frac{g_A}{m_\nu^2-m_\phi^2}\phi\,.
\end{align}

To first order in $g_A$, the residual mixing is proportional to the free equation of motion of $\phi$.
Up to boundary terms, it takes the form
\begin{align}
  S_{\rm mix}=\frac{g_A}{m_\nu^2-m_\phi^2}\int \dd^4x\sqrt{-g}\,\widetilde\chi_\nu(\Box-m_\phi^2)\phi\,.
\end{align}
The interaction at the hard vertex becomes
\begin{align}
  g_B\widetilde{\mathcal J}^{(\nu)}\chi_\nu = g_B\widetilde{\mathcal J}^{(\nu)} \left( \widetilde\chi_\nu +\frac{g_A}{m_\nu^2-m_\phi^2}\phi \right)\,. \label{redefinition}
\end{align}
At order $g_Ag_B$, we may use the free equation of motion of the external mode $(\Box-m_\phi^2)\phi=0$, with interaction corrections contributing only at higher orders. The term $S_{\rm mix}$ then gives no contribution, leaving only the contact interaction in \eqref{redefinition}. This interaction involves no $\chi_\nu$ exchange and hence no corresponding CC signal. The argument applies to each spectral component, so integrating over $\nu$ completes the proof of case (i). We thus see that oscillatory modes in the heavy sector need \textit{not} leave a CC signal, since their oscillations must also be transmitted to the external correlator and the covariant linear mixing at the soft vertex does not mediate this transfer.

\subsection{One-loop \texorpdfstring{$N$}{N}-gons}
We now turn to one-loop $N$-gons, whose heavy lines connect the soft vertex to several hard vertices. Let us consider the diagram in Fig.~\ref{fig:selection-rule}(c), with the soft external momentum $\bm k_L$ attached to the vertex at $\tau_N$. We choose the loop momentum $\bm q$ so that the two adjacent heavy propagators carry $\bm q$ and $\bm q+\bm k_L$, with their hard endpoints at $\tau_1$ and $\tau_{N-1}$, respectively. The $N$-gon contribution $\mathcal T_N$ to the correlator takes the schematic form\footnote{For simplicity, we present the argument for scalar fields. We expect it to carry over to spinning fields, whose spin structures only introduce additional tensor factors that should not alter the matching.}
\begin{align}
  \mathcal T_N &\sim \sum_{\mathsf a_1,\cdots,\mathsf a_N=\pm} \left[\prod_{v=1}^{N-1} \int_{-\infty}^0 \ii\mathsf a_v\dd\tau_v\,(-\tau_v)^{p_v}e^{\mathsf a_v\ii E_v\tau_v}\right]\nonumber\\
  &\times\int_{-\infty}^0 \ii\mathsf a_N\dd\tau_N\,(-\tau_N)^{p_N}e^{\mathsf a_N\ii k_L\tau_N}\times \mathcal {Q}\,, \label{TN}
\end{align}
where the loop integral is defined as
\begin{align}
  \mathcal{Q} &\equiv \int\frac{\dd^3\bm q}{(2\pi)^3}\,D_{\mathsf a_1\mathsf a_N}^{(\nu_1)}(q;\tau_1,\tau_N) D^{(\nu_N)}_{\mathsf a_N\mathsf a_{N-1}}(|\bm q+\bm k_L|;\tau_N,\tau_{N-1})\nonumber\\
  &\times \prod_{i=2}^{N-1} D^{(\nu_i)}_{\mathsf a_{i-1}\mathsf a_i}(|\bm q-\sum_{\ell=1}^{i-1}\bm k_\ell|;\tau_{i-1},\tau_i)\,. \label{QN}
\end{align}
Here $\mathsf a_i=\pm1$ label the SK branches, and $D_{\mathsf a_i\mathsf a_j}^{(\nu)}$ denotes the bulk-to-bulk propagator of the field with mass parameter $\nu$. The quantities $E_i$ and $\bm k_i$ denote the sum of external momentum magnitudes and the total external momentum entering vertex $i$, respectively. At the hard vertices $i=1,\cdots,N-1$, we have $E_i,k_i\gtrsim k_S$, whereas $E_N=k_N=k_L$ at the soft vertex. The integers $p_i$ account for all other time dependence.

To extract the CC signal, we first apply a directional cut to the two heavy propagators adjacent to the soft vertex. This replaces them by the Wightman components associated with the ordering $\tau_N<\tau_1,\tau_{N-1}$, while the other terms do not contribute to the CC signal \cite{Tong:2021wai,Ema:2024hkj,Qin:2026yuo}.

For the cut contribution, a CC signal can only come from the soft loop region $q\sim|\bm q+\bm k_L|\sim k_L\ll k_S$. At fixed $\tau_1$ and $\tau_{N-1}$, the corresponding mode arguments become small as $k_L\to0$. The two modes therefore admit superhorizon expansions at their hard endpoints, with oscillatory factors $(-q\tau_1)^{\pm\ii\nu_1}$ and $(-|\bm q+\bm k_L|\tau_{N-1})^{\pm\ii\nu_N}$, respectively.

We can now Taylor expand the hard subgraph in the soft momentum $\bm q$ and $\bm k_L$. At leading order, its dependence on the soft loop momentum drops out, leaving only the two soft internal lines in the loop integral. The hard subgraph can then be represented by a local ``pinched'' operator connecting the hard external legs, whose coefficient is fixed by matching the hard time integrals. This reduces the $N$-gon to a bubble diagram. At higher orders, the additional powers of soft momentum are reproduced by derivatives in the local operator, so the same matching procedure extends the reduction order by order in the soft expansion.\footnote{For illustration, let $\sigma_1$ and $\sigma_N$ denote the heavy fields carrying momentum $\bm q$ and $\bm p\equiv\bm q+\bm k_L$, respectively. A factor $q_i p_j$ can be generated by $(\nabla_i\sigma_1)(\nabla_j\sigma_N)$ at the hard vertex. Higher momentum monomials are reproduced by higher derivatives, with the indices contracted with those of the hard subgraph.}

Each matched bubble falls within case (i), for which we have already shown that the CC signal vanishes. The $N$-gon CC signal therefore vanishes at every order in the soft expansion. Schematically, the argument can be summarised by the diagrammatic relation \eqref{pinching}, and we put the detailed derivation in App.~\ref{sec:ngon}.
\begin{align}
\left[\vcenter{\hbox{\includegraphics{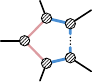}}}\right]_{\rm CC}
&\overset{\mathclap{\rm pinch}}{=}\quad \sum_r C_r
\left[\vcenter{\hbox{\includegraphics{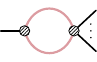}}}\right]_{\rm CC}\nonumber\\[0.8em]
&=\sum_{r,\nu} C_r\rho_r(\nu)
\left[ \vcenter{\hbox{\includegraphics{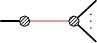}}}\right]_{\rm CC}=0\,. \label{pinching}
\end{align}

\subsection{Applications of the rule}
Over the past decade, CC physics has been used to explore a broad range of new physics scenarios including dark matter, Higgs inflation and leptogenesis, and many of the loop diagrams considered in these studies~\cite{Li:2020xwr,Chen:2016hrz,Ema:2023dxm,Lu:2019tjj,Cui:2021iie,Aoki:2020zbj} are covered by our exclusion rule. For example, the $W/Z$ bubbles in the cosmological Higgs collider~\cite{Lu:2019tjj} fall into case (i), while the neutrino triangle proposed as a probe of leptogenesis~\cite{Cui:2021iie} falls into case (ii), so that their oscillatory signals should vanish.

As such diagrams are difficult to evaluate analytically, earlier predictions were obtained by expanding the heavy field propagators at late times at both endpoints. Since the squeezed configuration does not validate such an expansion at the soft vertex, these predictions miss the contribution from the early time region. Previous results based on such expansions must therefore be revisited, including those used to calibrate the cosmological Higgs collider and to infer the parameters of leptogenesis. A nonzero signal in these processes requires an ingredient that evades the exclusion rule.

\section{Beyond the exclusion rule}
To generate a nonzero CC signal, at least one of the assumptions underlying the exclusion rule must be violated. A noncovariant soft vertex provides an immediate way to evade the rule. A nontrivial alternative is to keep the interaction local and dS covariant while breaking dS symmetry in the heavy sector, as we now illustrate with a chemical potential.

For concreteness, consider a heavy scalar $\sigma$ coupled to $\phi$ through the local covariant interaction
\begin{align}
  S_{\rm int}=-\frac{\lambda}{2}\int \dd^4x\sqrt{-g}\,\phi\sigma^2\,. \label{interaction}
\end{align}
The connected three-point function $\langle\phi\phi\phi\rangle$ receives a contribution at order $\lambda^3$ from three insertions of this interaction, forming the triangle in Fig.~\ref{fig:selection-rule}(d), the $N=3$ case of Fig.~\ref{fig:selection-rule}(c). Following our conventions, we take $\phi$ to be conformally coupled with $m_\phi^2=2H^2$ to simplify the calculation and avoid unnecessary subtleties.

Although a scalar does not admit a helical chemical potential, it can model the propagation of a spinning field with a chemical potential once the spin structures are omitted. These structures do not affect the oscillatory scaling of interest, so the scalar model allows us to show how the exclusion rule is evaded and carry out the calculation. For the mode function $u(k,\tau)$ of $\sigma$, we define the canonically normalised mode $v(k,\tau)=a(\tau)u(k,\tau)$ and take
\begin{align}
  \partial_\tau^2v + \left( k^2+\frac{2\kappa k}{\tau} +\frac{\nu^2+1/4}{\tau^2} \right)v = 0\,, \label{Whittaker}
\end{align}
where the term proportional to $\kappa$ describes the effect of the chemical potential and breaks dS boosts, while $\kappa=0$ recovers the covariant mode equation. The solution is given by a Whittaker-$W$ function, and we give the ingredients needed for the calculation in App.~\ref{sec:triangle}.

\begin{figure}[t]
  \centering
  \hspace{-1cm}\includegraphics[width=0.98\columnwidth]{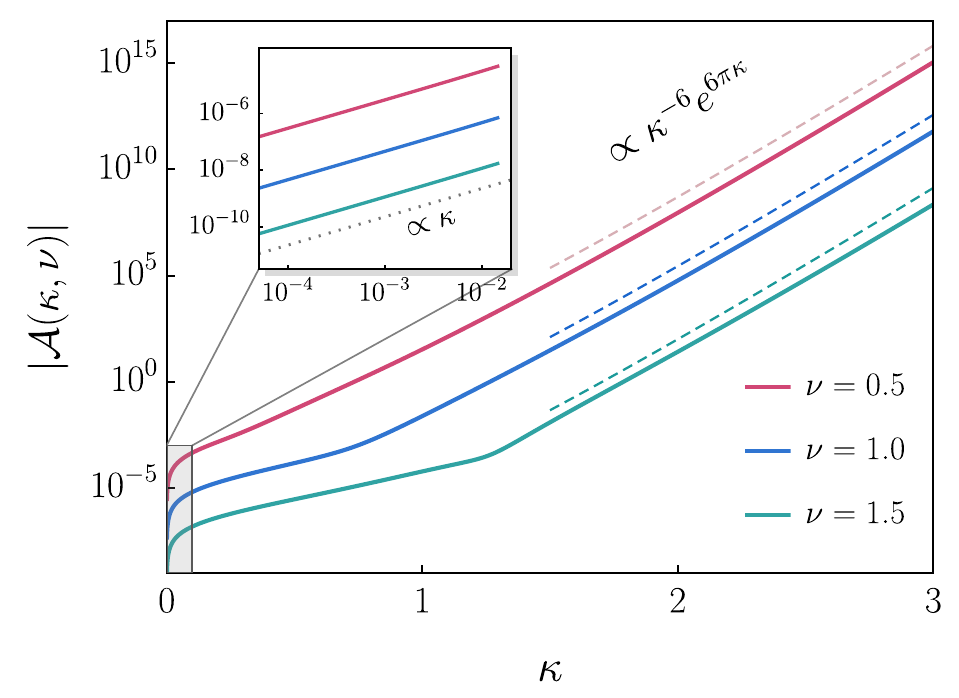}
  \caption{CC signals from a loop with a chemical potential term for different fixed masses. The dashed curves indicate the scaling $\kappa^{-6}e^{6\pi\kappa}$ at large $\kappa$. The inset resolves the behaviour at small $\kappa$, where the amplitude approaches zero linearly with the chemical potential.}
  \label{fig:chemical-potential-signal}
\end{figure}

Triangle loops are notoriously difficult to evaluate analytically, and the presence of a chemical potential further complicates the calculation. The pinching procedure~\cite{Qin:2026yuo} used in the proof above reduces the triangle to bubble integrals order by order in the soft expansion. These integrals can then be evaluated using Mellin-Barnes (MB) representations of the Whittaker propagators. To our knowledge, this provides the first analytical result for the CC signal from such a triangle with a chemical potential. Details of the calculation are given in App.~\ref{sec:triangle}, while a more extensive treatment of loop CC signals with a chemical potential will be presented in a forthcoming companion paper.

Let us first define the dimensionless bispectrum $B$ by
\begin{align}
  \langle\phi\phi\phi\rangle' = \frac{\lambda^3\tau_f^3k_3} {(k_1+k_2)^4} B(k_1,k_2,k_3)\,, \label{bispectrum}
\end{align}
where $\tau_f$ denotes the end of inflation and the prime indicates that the momentum conservation delta function has been removed. Since each conformally coupled external mode scales as $\tau$ at late times, the three-point function carries an overall factor of $\tau_f^3$, which we have factored out in defining the bispectrum. The CC signal is given by the oscillatory part of $B(k_1,k_2,k_3)$ in the single soft limit $k_3\equiv k_L\ll k_1,k_2$. At leading order, it can be formally expressed as
\begin{align}
  B_{\rm CC} \simeq \mathcal A(\kappa,\nu) \left(\frac{k_L}{k_S}\right)^{2\ii\nu} +\text{c.c.}\,,
\end{align}
where $k_S\equiv k_1+k_2$ and the complex coefficient $\mathcal A(\kappa,\nu)$ determines the amplitude and phase of the signal. The closed form of $\mathcal A$ is lengthy, so we give it in App.~\ref{sec:triangle} and omit it here. By breaking dS symmetry in the heavy sector, the chemical potential evades the exclusion rule and produces the nonzero amplitudes shown in Fig.~\ref{fig:chemical-potential-signal} for different fixed masses. The rapid growth at large $\kappa$ reflects the enhancement of each heavy propagator by a factor $e^{2\pi\kappa}$, giving an overall factor $e^{6\pi\kappa}$ for the triangle. We have verified numerically that the amplitude follows the scaling $\kappa^{-6}e^{6\pi\kappa}$, as indicated by the dashed curves. The limit of zero chemical potential provides a further nontrivial check of the result. As shown in the inset, the amplitude approaches zero linearly with $\kappa$. At $\kappa=0$, dS symmetry is restored and the exclusion rule requires $\mathcal A(0,\nu)=0$, which we have verified explicitly. Further details can be found in App.~\ref{sec:triangle}.

\section{Summary}
The late-time oscillations of heavy field modes can be imprinted on correlators through interactions with the external field. One might therefore expect a nonzero coupling to these modes to be sufficient to generate oscillations in momentum ratios. Our results show that the corresponding CC signal can nevertheless be entirely absent.

In this \textit{letter}, we have derived an exclusion rule that identifies when this silence occurs. The rule applies whenever a dS invariant heavy sector couples covariantly to a single external field at one vertex. For diagrams that connect this vertex to the other vertex through a two-point function of the heavy sector, we proved that the CC signal vanishes to all loop orders and in any kinematic configuration. We then extended the rule to one-loop polygons, for which the CC signal vanishes at every order in the soft expansion. In both classes, the result holds regardless of the interactions at other vertices.

We also showed how breaking dS symmetry in the heavy sector can evade the exclusion rule. In a scalar triangle model with Whittaker propagation motivated by helical chemical potentials, we obtained the first analytical result for the leading CC signal. The signal is generically nonzero and vanishes linearly as the chemical potential is switched off, recovering the null result required by the rule in the dS invariant limit.

Our results suggest several directions for further investigation. For instance, the vanishing of the triangle signal at zero chemical potential can be understood from the late-time boundary, since all interactions then respect dS symmetry and conformal invariance fixes the three-point function up to an overall coefficient, leaving no room for a CC signal. The exclusion rule goes beyond this case, covering higher-point correlators and arbitrary interactions at the hard vertices. In particular, a box diagram with dS covariant interactions contributes no CC signal to the trispectrum in the soft limit of one external momentum, although conformal invariance does not fix this four-point function. It would be interesting to understand these vanishing signals from the boundary perspective as well. Another natural direction is to extend the rule to higher-loop diagrams with arbitrary hard subgraphs attached to the soft vertex. In addition, spin deserves a detailed treatment for both the external and the heavy fields, which we will partly address in a forthcoming companion paper. Finally, it would be valuable to revisit the phenomenological predictions of previously studied loop processes in light of the exclusion rule.

\vspace{5pt}
\paragraph*{Acknowledgements}
We thank Chengcheng Han, Enrico Pajer, Dong-Gang Wang, Yi Wang, and Zhong-Zhi Xianyu for useful discussions.
S.A. is supported by the Japan Science and Technology Agency (JST) as part of Adopting Sustainable Partnerships for Innovative Research Ecosystem (ASPIRE), grant JPMJAP2318, and by JSPS KAKENHI Grant Number 26K07078. Z.Q. is supported by the China Scholarship Council (CSC). M.Y. is supported by IBS under the project code, IBS-R018-D3, and by JSPS Grant-in-Aid for Scientific Research Number JP23K20843. Y.Z. is supported by IBS under the project code, IBS-R018-D3. The authors used ChatGPT (OpenAI, GPT-5.6 Sol and GPT-6 Astra) and Claude (Anthropic, Fable 5.1) to assist in checking derivations, identifying possible inconsistencies, and improving the presentation of the manuscript. All AI-assisted outputs were independently checked by the authors, who take full responsibility for the content.

\bibliographystyle{utphys}
\bibliography{references}

\providecommand{\href}[2]{#2}\begingroup\raggedright\begin{thebibliography}{10}

\bibitem{Chen:2009zp}
X.~Chen and Y.~Wang, ``{Quasi-Single Field Inflation and Non-Gaussianities},''
  \href{http://dx.doi.org/10.1088/1475-7516/2010/04/027}{{\em JCAP} {\bfseries
  1004} (2010) 027},
\href{http://arxiv.org/abs/0911.3380}{{\ttfamily arXiv:0911.3380 [hep-th]}}.

\bibitem{Baumann:2011nk}
D.~Baumann and D.~Green, ``{Signatures of Supersymmetry from the Early
  Universe},'' \href{http://dx.doi.org/10.1103/PhysRevD.85.103520}{{\em Phys.
  Rev.} {\bfseries D85} (2012) 103520},
\href{http://arxiv.org/abs/1109.0292}{{\ttfamily arXiv:1109.0292 [hep-th]}}.

\bibitem{Noumi:2012vr}
T.~Noumi, M.~Yamaguchi, and D.~Yokoyama, ``{Effective field theory approach to
  quasi-single field inflation and effects of heavy fields},''
  \href{http://dx.doi.org/10.1007/JHEP06(2013)051}{{\em JHEP} {\bfseries 06}
  (2013) 051}, \href{http://arxiv.org/abs/1211.1624}{{\ttfamily arXiv:1211.1624
  [hep-th]}}.

\bibitem{Arkani-Hamed:2015bza}
N.~Arkani-Hamed and J.~Maldacena, ``{Cosmological Collider Physics},''
\href{http://arxiv.org/abs/1503.08043}{{\ttfamily arXiv:1503.08043 [hep-th]}}.

\bibitem{Arkani-Hamed:2018kmz}
N.~Arkani-Hamed, D.~Baumann, H.~Lee, and G.~L. Pimentel, ``{The Cosmological
  Bootstrap: Inflationary Correlators from Symmetries and Singularities},''
  \href{http://dx.doi.org/10.1007/JHEP04(2020)105}{{\em JHEP} {\bfseries 04}
  (2020) 105}, \href{http://arxiv.org/abs/1811.00024}{{\ttfamily
  arXiv:1811.00024 [hep-th]}}.

\bibitem{Baumann:2019oyu}
D.~Baumann, C.~Duaso~Pueyo, A.~Joyce, H.~Lee, and G.~L. Pimentel, ``{The
  cosmological bootstrap: weight-shifting operators and scalar seeds},''
  \href{http://dx.doi.org/10.1007/JHEP12(2020)204}{{\em JHEP} {\bfseries 12}
  (2020) 204}, \href{http://arxiv.org/abs/1910.14051}{{\ttfamily
  arXiv:1910.14051 [hep-th]}}.

\bibitem{Sleight:2019mgd}
C.~Sleight, ``{A Mellin Space Approach to Cosmological Correlators},''
  \href{http://dx.doi.org/10.1007/JHEP01(2020)090}{{\em JHEP} {\bfseries 01}
  (2020) 090}, \href{http://arxiv.org/abs/1906.12302}{{\ttfamily
  arXiv:1906.12302 [hep-th]}}.

\bibitem{Sleight:2019hfp}
C.~Sleight and M.~Taronna, ``{Bootstrapping Inflationary Correlators in Mellin
  Space},'' \href{http://dx.doi.org/10.1007/JHEP02(2020)098}{{\em JHEP}
  {\bfseries 02} (2020) 098}, \href{http://arxiv.org/abs/1907.01143}{{\ttfamily
  arXiv:1907.01143 [hep-th]}}.

\bibitem{Sleight:2020obc}
C.~Sleight and M.~Taronna, ``{From AdS to dS exchanges: Spectral
  representation, Mellin amplitudes, and crossing},''
  \href{http://dx.doi.org/10.1103/PhysRevD.104.L081902}{{\em Phys. Rev. D}
  {\bfseries 104} no.~8, (2021) L081902},
  \href{http://arxiv.org/abs/2007.09993}{{\ttfamily arXiv:2007.09993
  [hep-th]}}.

\bibitem{Pajer:2020wxk}
E.~Pajer, ``{Building a Boostless Bootstrap for the Bispectrum},''
  \href{http://dx.doi.org/10.1088/1475-7516/2021/01/023}{{\em JCAP} {\bfseries
  01} (2021) 023}, \href{http://arxiv.org/abs/2010.12818}{{\ttfamily
  arXiv:2010.12818 [hep-th]}}.

\bibitem{Jazayeri:2021fvk}
S.~Jazayeri, E.~Pajer, and D.~Stefanyszyn, ``{From locality and unitarity to
  cosmological correlators},''
  \href{http://dx.doi.org/10.1007/JHEP10(2021)065}{{\em JHEP} {\bfseries 10}
  (2021) 065}, \href{http://arxiv.org/abs/2103.08649}{{\ttfamily
  arXiv:2103.08649 [hep-th]}}.

\bibitem{Goodhew:2020hob}
H.~Goodhew, S.~Jazayeri, and E.~Pajer, ``{The Cosmological Optical Theorem},''
  \href{http://dx.doi.org/10.1088/1475-7516/2021/04/021}{{\em JCAP} {\bfseries
  04} (2021) 021}, \href{http://arxiv.org/abs/2009.02898}{{\ttfamily
  arXiv:2009.02898 [hep-th]}}.

\bibitem{Melville:2021lst}
S.~Melville and E.~Pajer, ``{Cosmological Cutting Rules},''
  \href{http://dx.doi.org/10.1007/JHEP05(2021)249}{{\em JHEP} {\bfseries 05}
  (2021) 249}, \href{http://arxiv.org/abs/2103.09832}{{\ttfamily
  arXiv:2103.09832 [hep-th]}}.

\bibitem{Pimentel:2022fsc}
G.~L. Pimentel and D.-G. Wang, ``{Boostless cosmological collider bootstrap},''
  \href{http://dx.doi.org/10.1007/JHEP10(2022)177}{{\em JHEP} {\bfseries 10}
  (2022) 177}, \href{http://arxiv.org/abs/2205.00013}{{\ttfamily
  arXiv:2205.00013 [hep-th]}}.

\bibitem{Jazayeri:2022kjy}
S.~Jazayeri and S.~Renaux-Petel, ``{Cosmological bootstrap in slow motion},''
  \href{http://dx.doi.org/10.1007/JHEP12(2022)137}{{\em JHEP} {\bfseries 12}
  (2022) 137}, \href{http://arxiv.org/abs/2205.10340}{{\ttfamily
  arXiv:2205.10340 [hep-th]}}.

\bibitem{Wang:2022eop}
D.-G. Wang, G.~L. Pimentel, and A.~Ach\'ucarro, ``{Bootstrapping multi-field
  inflation: non-Gaussianities from light scalars revisited},''
  \href{http://dx.doi.org/10.1088/1475-7516/2023/05/043}{{\em JCAP} {\bfseries
  05} (2023) 043}, \href{http://arxiv.org/abs/2212.14035}{{\ttfamily
  arXiv:2212.14035 [astro-ph.CO]}}.

\bibitem{Qin:2022fbv}
Z.~Qin and Z.-Z. Xianyu, ``{Helical inflation correlators: partial
  Mellin-Barnes and bootstrap equations},''
  \href{http://dx.doi.org/10.1007/JHEP04(2023)059}{{\em JHEP} {\bfseries 04}
  (2023) 059}, \href{http://arxiv.org/abs/2208.13790}{{\ttfamily
  arXiv:2208.13790 [hep-th]}}.

\bibitem{Qin:2023ejc}
Z.~Qin and Z.-Z. Xianyu, ``{Closed-form formulae for inflation correlators},''
  \href{http://dx.doi.org/10.1007/JHEP07(2023)001}{{\em JHEP} {\bfseries 07}
  (2023) 001}, \href{http://arxiv.org/abs/2301.07047}{{\ttfamily
  arXiv:2301.07047 [hep-th]}}.

\bibitem{De:2023xue}
S.~De and A.~Pokraka, ``{Cosmology meets cohomology},''
  \href{http://dx.doi.org/10.1007/JHEP03(2024)156}{{\em JHEP} {\bfseries 03}
  (2024) 156}, \href{http://arxiv.org/abs/2308.03753}{{\ttfamily
  arXiv:2308.03753 [hep-th]}}.

\bibitem{Arkani-Hamed:2023bsv}
N.~Arkani-Hamed, D.~Baumann, A.~Hillman, A.~Joyce, H.~Lee, and G.~L. Pimentel,
  ``{Kinematic Flow and the Emergence of Time},''
  \href{http://dx.doi.org/10.1103/dsjm-tckw}{{\em Phys. Rev. Lett.} {\bfseries
  135} no.~3, (2025) 031602}, \href{http://arxiv.org/abs/2312.05300}{{\ttfamily
  arXiv:2312.05300 [hep-th]}}.

\bibitem{Stefanyszyn:2023qov}
D.~Stefanyszyn, X.~Tong, and Y.~Zhu, ``{Cosmological correlators through the
  looking glass: reality, parity, and factorisation},''
  \href{http://dx.doi.org/10.1007/JHEP05(2024)196}{{\em JHEP} {\bfseries 05}
  (2024) 196}, \href{http://arxiv.org/abs/2309.07769}{{\ttfamily
  arXiv:2309.07769 [hep-th]}}.

\bibitem{Aoki:2024uyi}
S.~Aoki, L.~Pinol, F.~Sano, M.~Yamaguchi, and Y.~Zhu, ``{Cosmological
  correlators with double massive exchanges: bootstrap equation and
  phenomenology},'' \href{http://dx.doi.org/10.1007/JHEP09(2024)176}{{\em JHEP}
  {\bfseries 09} (2024) 176}, \href{http://arxiv.org/abs/2404.09547}{{\ttfamily
  arXiv:2404.09547 [hep-th]}}.

\bibitem{Liu:2024str}
H.~Liu and Z.-Z. Xianyu, ``{Massive inflationary amplitudes: differential
  equations and complete solutions for general trees},''
  \href{http://dx.doi.org/10.1007/JHEP09(2025)183}{{\em JHEP} {\bfseries 09}
  (2025) 183}, \href{http://arxiv.org/abs/2412.07843}{{\ttfamily
  arXiv:2412.07843 [hep-th]}}.

\bibitem{Qin:2025xct}
Z.~Qin, S.~Renaux-Petel, X.~Tong, D.~Werth, and Y.~Zhu, ``{The exact and
  approximate tales of boost-breaking cosmological correlators},''
  \href{http://dx.doi.org/10.1088/1475-7516/2025/09/058}{{\em JCAP} {\bfseries
  09} (2025) 058}, \href{http://arxiv.org/abs/2506.01555}{{\ttfamily
  arXiv:2506.01555 [hep-th]}}.

\bibitem{Werth:2024mjg}
D.~Werth, ``{Spectral representation of cosmological correlators},''
  \href{http://dx.doi.org/10.1007/JHEP12(2024)017}{{\em JHEP} {\bfseries 12}
  (2024) 017}, \href{http://arxiv.org/abs/2409.02072}{{\ttfamily
  arXiv:2409.02072 [hep-th]}}.

\bibitem{Chen:2024glu}
J.~Chen, B.~Feng, and Y.-X. Tao, ``{Multivariate hypergeometric solutions of
  cosmological (dS) correlators by d log-form differential equations},''
  \href{http://dx.doi.org/10.1007/JHEP03(2025)075}{{\em JHEP} {\bfseries 03}
  (2025) 075}, \href{http://arxiv.org/abs/2411.03088}{{\ttfamily
  arXiv:2411.03088 [hep-th]}}.

\bibitem{Stefanyszyn:2024msm}
D.~Stefanyszyn, X.~Tong, and Y.~Zhu, ``{There and Back Again: Mapping and
  Factorizing Cosmological Observables},''
  \href{http://dx.doi.org/10.1103/PhysRevLett.133.221501}{{\em Phys. Rev.
  Lett.} {\bfseries 133} no.~22, (2024) 221501},
  \href{http://arxiv.org/abs/2406.00099}{{\ttfamily arXiv:2406.00099
  [hep-th]}}.

\bibitem{Pimentel:2025rds}
G.~L. Pimentel and C.~Yang, ``{Strongly coupled sectors in inflation: gapless
  theories and unparticles},''
  \href{http://dx.doi.org/10.1007/JHEP04(2026)146}{{\em JHEP} {\bfseries 04}
  (2026) 146}, \href{http://arxiv.org/abs/2503.17840}{{\ttfamily
  arXiv:2503.17840 [hep-th]}}.

\bibitem{Cheung:2025dmc}
T.~Cheung and D.~Stefanyszyn, ``{Massive spinning fields during inflation:
  Feynman rules and correlator comparison},''
  \href{http://dx.doi.org/10.1088/1475-7516/2026/05/022}{{\em JCAP} {\bfseries
  05} (2026) 022}, \href{http://arxiv.org/abs/2509.08888}{{\ttfamily
  arXiv:2509.08888 [hep-th]}}.

\bibitem{Chowdhury:2025ohm}
C.~Chowdhury, A.~Lipstein, J.~Marshall, J.~Mei, and I.~Sachs, ``{Cosmological
  dressing rules},'' \href{http://dx.doi.org/10.1007/JHEP03(2026)076}{{\em
  JHEP} {\bfseries 03} (2026) 076},
  \href{http://arxiv.org/abs/2503.10598}{{\ttfamily arXiv:2503.10598
  [hep-th]}}.

\bibitem{Jazayeri:2025vlv}
S.~Jazayeri, X.~Tong, and Y.~Zhu, ``{Every wrinkle carries a memory: an
  integro-differential bootstrap for features in cosmological correlators},''
  \href{http://dx.doi.org/10.1007/JHEP06(2026)022}{{\em JHEP} {\bfseries 06}
  (2026) 022}, \href{http://arxiv.org/abs/2511.00152}{{\ttfamily
  arXiv:2511.00152 [hep-th]}}.

\bibitem{Cheung:2026wrz}
T.~Cheung, C.~Duaso-Pueyo, D.~Stefanyszyn, and D.-G. Wang, ``{General Spin-1
  Dynamics During Inflation: Microcausality Constraints and Cosmological
  Collider Signals},'' \href{http://arxiv.org/abs/2610.10693}{{\ttfamily
  arXiv:2610.10693 [hep-th]}}.

\bibitem{Huenupi:2026abj}
J.~Huenupi, C.~Mu{\~n}oz, G.~A. Palma, and S.~Sypsas, ``{Pushing the Primordial
  Frontier: Exact Linear Solutions in Multifield Inflation},''
  \href{http://arxiv.org/abs/2606.18248}{{\ttfamily arXiv:2606.18248
  [astro-ph.CO]}}.

\bibitem{Arundine:2026myr}
M.~Arundine and G.~L. Pimentel, ``{Cosmological Collider in the
  Grassmannian},'' \href{http://arxiv.org/abs/2605.21581}{{\ttfamily
  arXiv:2605.21581 [hep-th]}}.

\bibitem{Ferreira:2026awc}
M.~C. Ferreira and F.~T. Falciano, ``{Bootstrapping a Higher-Derivative
  Cosmological Collider: Borel-Laplace Resummation and Stokes Phenomena},''
  \href{http://arxiv.org/abs/2610.42020}{{\ttfamily arXiv:2610.42020
  [hep-th]}}.

\bibitem{Belrhali:2026ktb}
N.~Belrhali, A.~Poisson, S.~Renaux-Petel, and D.~Werth, ``{De Sitter Momentum
  Space},'' \href{http://arxiv.org/abs/2601.15228}{{\ttfamily arXiv:2601.15228
  [hep-th]}}.

\bibitem{Belrhali:2026ygh}
N.~Belrhali, A.~Poisson, and S.~Renaux-Petel, ``{Laplace Space for Cosmological
  Correlators},'' \href{http://arxiv.org/abs/2606.27309}{{\ttfamily
  arXiv:2606.27309 [hep-th]}}.

\bibitem{Grafe:2026avi}
J.~Gr{\"a}fe and D.~Werth, ``{All Tree-Level Massive Cosmological Correlators
  via Spectral Gluing},'' \href{http://arxiv.org/abs/2607.18223}{{\ttfamily
  arXiv:2607.18223 [hep-th]}}.

\bibitem{Arkani-Hamed:2017fdk}
N.~Arkani-Hamed, P.~Benincasa, and A.~Postnikov, ``{Cosmological Polytopes and
  the Wavefunction of the Universe},''
  \href{http://arxiv.org/abs/1709.02813}{{\ttfamily arXiv:1709.02813
  [hep-th]}}.

\bibitem{Cabass:2024wob}
G.~Cabass, O.~H.~E. Philcox, M.~M. Ivanov, K.~Akitsu, S.-F. Chen,
  M.~Simonovi{\'c}, and M.~Zaldarriaga, ``{BOSS constraints on massive
  particles during inflation: The cosmological collider in action},''
  \href{http://dx.doi.org/10.1103/PhysRevD.111.063510}{{\em Phys. Rev. D}
  {\bfseries 111} no.~6, (2025) 063510},
  \href{http://arxiv.org/abs/2404.01894}{{\ttfamily arXiv:2404.01894
  [astro-ph.CO]}}.

\bibitem{Goldstein:2024bky}
S.~Goldstein, O.~H.~E. Philcox, J.~C. Hill, and L.~Hui, ``{Intermediate
  mass-range particles from small scales: Nonperturbative techniques for
  cosmological collider physics from large-scale structure surveys},''
  \href{http://dx.doi.org/10.1103/PhysRevD.110.083516}{{\em Phys. Rev. D}
  {\bfseries 110} no.~8, (2024) 083516},
  \href{http://arxiv.org/abs/2407.08731}{{\ttfamily arXiv:2407.08731
  [astro-ph.CO]}}.

\bibitem{Sohn:2024xzd}
W.~Sohn, D.-G. Wang, J.~R. Fergusson, and E.~P.~S. Shellard, ``{Searching for
  cosmological collider in the Planck CMB data},''
  \href{http://dx.doi.org/10.1088/1475-7516/2024/09/016}{{\em JCAP} {\bfseries
  09} (2024) 016}, \href{http://arxiv.org/abs/2404.07203}{{\ttfamily
  arXiv:2404.07203 [astro-ph.CO]}}.

\bibitem{Suman:2025vuf}
P.~Suman, D.-G. Wang, W.~Sohn, J.~R. Fergusson, and E.~P.~S. Shellard, ``{How
  Significant are Cosmological Collider Signals in the Planck Data?},''
  \href{http://arxiv.org/abs/2511.17500}{{\ttfamily arXiv:2511.17500
  [astro-ph.CO]}}.

\bibitem{Suman:2025tpv}
P.~Suman, D.-G. Wang, W.~Sohn, J.~R. Fergusson, and E.~P.~S. Shellard,
  ``{Searching for Cosmological Collider in the Planck CMB Data II: collider
  templates and Modal analysis},''
  \href{http://arxiv.org/abs/2512.22085}{{\ttfamily arXiv:2512.22085
  [astro-ph.CO]}}.

\bibitem{Philcox:2025wts}
O.~H.~E. Philcox, ``{Searching for inflationary physics with the CMB
  trispectrum. III. Constraints from Planck},''
  \href{http://dx.doi.org/10.1103/y81z-g7th}{{\em Phys. Rev. D} {\bfseries 111}
  no.~12, (2025) 123534}, \href{http://arxiv.org/abs/2502.06931}{{\ttfamily
  arXiv:2502.06931 [astro-ph.CO]}}.

\bibitem{Anbajagane:2025uro}
D.~Anbajagane and H.~Lee, ``{Primordial physics in the nonlinear Universe:
  Mapping cosmological collider models to weak-lensing observables},''
  \href{http://dx.doi.org/10.1088/1475-7516/2026/05/076}{{\em JCAP} {\bfseries
  05} (2026) 076}, \href{http://arxiv.org/abs/2509.02693}{{\ttfamily
  arXiv:2509.02693 [astro-ph.CO]}}.

\bibitem{Green:2026yev}
D.~Green, J.~Han, and B.~Wallisch, ``{Extending the Cosmological Collider: New
  Scaling Regimes and Constraints from BOSS},''
  \href{http://arxiv.org/abs/2602.12232}{{\ttfamily arXiv:2602.12232
  [astro-ph.CO]}}.

\bibitem{Kumar:2026ogn}
S.~Kumar, Q.~Lu, Z.-Z. Xianyu, and Y.~Zhang, ``{Cosmological Collider Searches
  beyond the Hubble Scale with Planck Data},''
  \href{http://arxiv.org/abs/2603.15728}{{\ttfamily arXiv:2603.15728
  [hep-ph]}}.

\bibitem{Kumar:2026dih}
S.~Kumar, Q.~Lu, Z.-Z. Xianyu, and Y.~Zhang, ``{Scalars at the Cosmological
  Collider: Full Shapes of Tree Diagrams and Bispectrum Searches using Planck
  Data},'' \href{http://arxiv.org/abs/2604.07434}{{\ttfamily arXiv:2604.07434
  [hep-ph]}}.

\bibitem{Philcox:2026tjj}
O.~H.~E. Philcox, ``{Dissecting the Scalar Cosmological Collider with the
  Cosmic Microwave Background},''
  \href{http://arxiv.org/abs/2607.18369}{{\ttfamily arXiv:2607.18369
  [astro-ph.CO]}}.

\bibitem{Xianyu:2022jwk}
Z.-Z. Xianyu and H.~Zhang, ``{Bootstrapping one-loop inflation correlators with
  the spectral decomposition},''
  \href{http://dx.doi.org/10.1007/JHEP04(2023)103}{{\em JHEP} {\bfseries 04}
  (2023) 103}, \href{http://arxiv.org/abs/2211.03810}{{\ttfamily
  arXiv:2211.03810 [hep-th]}}.

\bibitem{Salcedo:2022aal}
S.~A. Salcedo, M.~H.~G. Lee, S.~Melville, and E.~Pajer, ``{The Analytic
  Wavefunction},'' \href{http://dx.doi.org/10.1007/JHEP06(2023)020}{{\em JHEP}
  {\bfseries 06} (2023) 020}, \href{http://arxiv.org/abs/2212.08009}{{\ttfamily
  arXiv:2212.08009 [hep-th]}}.

\bibitem{Lee:2023jby}
M.~H.~G. Lee, C.~McCulloch, and E.~Pajer, ``{Leading loops in cosmological
  correlators},'' \href{http://dx.doi.org/10.1007/JHEP11(2023)038}{{\em JHEP}
  {\bfseries 11} (2023) 038}, \href{http://arxiv.org/abs/2305.11228}{{\ttfamily
  arXiv:2305.11228 [hep-th]}}.

\bibitem{Qin:2023bjk}
Z.~Qin and Z.-Z. Xianyu, ``{Inflation correlators at the one-loop order:
  nonanalyticity, factorization, cutting rule, and OPE},''
  \href{http://dx.doi.org/10.1007/JHEP09(2023)116}{{\em JHEP} {\bfseries 09}
  (2023) 116}, \href{http://arxiv.org/abs/2304.13295}{{\ttfamily
  arXiv:2304.13295 [hep-th]}}.

\bibitem{Cespedes:2023aal}
S.~C{\'e}spedes, A.-C. Davis, and D.-G. Wang, ``{On the IR divergences in de
  Sitter space: loops, resummation and the semi-classical wavefunction},''
  \href{http://dx.doi.org/10.1007/JHEP04(2024)004}{{\em JHEP} {\bfseries 04}
  (2024) 004}, \href{http://arxiv.org/abs/2311.17990}{{\ttfamily
  arXiv:2311.17990 [hep-th]}}.

\bibitem{AguiSalcedo:2023nds}
S.~Agui~Salcedo and S.~Melville, ``{The cosmological tree theorem},''
  \href{http://dx.doi.org/10.1007/JHEP12(2023)076}{{\em JHEP} {\bfseries 12}
  (2023) 076}, \href{http://arxiv.org/abs/2308.00680}{{\ttfamily
  arXiv:2308.00680 [hep-th]}}.

\bibitem{Qin:2024gtr}
Z.~Qin, ``{Cosmological correlators at the loop level},''
  \href{http://dx.doi.org/10.1007/JHEP03(2025)051}{{\em JHEP} {\bfseries 03}
  (2025) 051}, \href{http://arxiv.org/abs/2411.13636}{{\ttfamily
  arXiv:2411.13636 [hep-th]}}.

\bibitem{Bhowmick:2025mxh}
S.~Bhowmick, M.~H.~G. Lee, D.~Ghosh, and F.~Ullah, ``{Singularities in
  cosmological loop correlators},''
  \href{http://dx.doi.org/10.1007/JHEP02(2026)116}{{\em JHEP} {\bfseries 02}
  (2026) 116}, \href{http://arxiv.org/abs/2503.21880}{{\ttfamily
  arXiv:2503.21880 [hep-th]}}.

\bibitem{Jain:2025maa}
D.~Jain, E.~Pajer, and X.~Tong, ``{Unitary and Analytic Renormalisation of
  Cosmological Correlators},''
  \href{http://dx.doi.org/10.1007/JHEP03(2026)225}{{\em JHEP} {\bfseries 03}
  (2026) 225}, \href{http://arxiv.org/abs/2509.02696}{{\ttfamily
  arXiv:2509.02696 [hep-th]}}.

\bibitem{Cespedes:2025ple}
S.~Cespedes, Z.~Qin, and D.-G. Wang,
  ``{{\ensuremath{\lambda}}{\ensuremath{\phi}}$^{4}$ as an effective theory in
  de Sitter},'' \href{http://dx.doi.org/10.1007/JHEP05(2026)143}{{\em JHEP}
  {\bfseries 05} (2026) 143}, \href{http://arxiv.org/abs/2510.25826}{{\ttfamily
  arXiv:2510.25826 [hep-th]}}.

\bibitem{Baumann:2025qjx}
D.~Baumann, H.~Goodhew, A.~Joyce, H.~Lee, G.~L. Pimentel, and T.~Westerdijk,
  ``{Geometry of kinematic flow},''
  \href{http://dx.doi.org/10.1007/JHEP05(2026)211}{{\em JHEP} {\bfseries 05}
  (2026) 211}, \href{http://arxiv.org/abs/2504.14890}{{\ttfamily
  arXiv:2504.14890 [hep-th]}}.

\bibitem{Pimentel:2026kqc}
G.~L. Pimentel and T.~Westerdijk, ``{On Cosmological Correlators at One
  Loop},'' \href{http://arxiv.org/abs/2601.00952}{{\ttfamily arXiv:2601.00952
  [hep-th]}}.

\bibitem{Grafe:2026qsm}
J.~Gr{\"a}fe and I.~Sachs, ``{Split representations and bubble resummation for
  massive de Sitter correlators},''
  \href{http://dx.doi.org/10.1007/JHEP07(2026)074}{{\em JHEP} {\bfseries 07}
  (2026) 074}, \href{http://arxiv.org/abs/2602.09977}{{\ttfamily
  arXiv:2602.09977 [hep-th]}}.

\bibitem{Farren:2026hao}
A.~Farren, C.~McCulloch, E.~Pajer, and X.~Tong, ``{All-Loop Renormalization and
  the Phase of the de Sitter Wavefunction},''
  \href{http://arxiv.org/abs/2603.08794}{{\ttfamily arXiv:2603.08794
  [hep-th]}}.

\bibitem{Westerdijk:2025ywh}
T.~Westerdijk and C.~Yang, ``{Bananas are unparticles: differential equations
  and cosmological bootstrap},''
  \href{http://dx.doi.org/10.1007/JHEP09(2025)089}{{\em JHEP} {\bfseries 09}
  (2025) 089}, \href{http://arxiv.org/abs/2503.08775}{{\ttfamily
  arXiv:2503.08775 [hep-th]}}.

\bibitem{Baumann:2026atn}
D.~Baumann, A.~Joyce, H.~Lee, and K.~Salehi~Vaziri, ``{Differential Equations
  for Massive Correlators},'' \href{http://arxiv.org/abs/2604.08658}{{\ttfamily
  arXiv:2604.08658 [hep-th]}}.

\bibitem{Chen:2026dqp}
J.~Chen, B.~Feng, Z.~Qin, and Y.-X. Tao, ``{Loop integrals in de Sitter
  spacetime: The parity-split IBP system and $\mathrm{d}\log$-form differential
  equations},'' \href{http://arxiv.org/abs/2604.14549}{{\ttfamily
  arXiv:2604.14549 [hep-th]}}.

\bibitem{Aoki:2026vbc}
S.~Aoki, Z.~Qin, M.~Yamaguchi, and Y.~Zhu, ``{Fermionic Bubble Loop in
  Cosmological Collider Revisited: Exact signals from spectral and
  Mellin-Barnes methods},'' \href{http://arxiv.org/abs/2605.28054}{{\ttfamily
  arXiv:2605.28054 [hep-th]}}.

\bibitem{Liu:2026jzn}
H.~Liu, Z.~Qin, J.~Wu, Z.-Z. Xianyu, and H.~Zhang, ``{On-Shell Bootstrap of
  Loop Inflation Correlators with Spectral Dispersion},''
  \href{http://arxiv.org/abs/2606.02686}{{\ttfamily arXiv:2606.02686
  [hep-th]}}.

\bibitem{Qin:2026yuo}
Z.~Qin and Z.-Z. Xianyu, ``{Cosmological Collider Signals From a Triangle
  Loop},'' \href{http://arxiv.org/abs/2609.09288}{{\ttfamily arXiv:2609.09288
  [hep-th]}}.

\bibitem{Borinsky:2026roj}
M.~Borinsky, A.~Herderschee, and Q.~Lu, ``{On the Numerical Integration of
  One-Loop Cosmological Collider Signals},''
  \href{http://arxiv.org/abs/2609.10673}{{\ttfamily arXiv:2609.10673
  [hep-ph]}}.

\bibitem{Chen:2016uwp}
X.~Chen, Y.~Wang, and Z.-Z. Xianyu, ``{Standard Model Background of the
  Cosmological Collider},''
  \href{http://dx.doi.org/10.1103/PhysRevLett.118.261302}{{\em Phys. Rev.
  Lett.} {\bfseries 118} no.~26, (2017) 261302},
\href{http://arxiv.org/abs/1610.06597}{{\ttfamily arXiv:1610.06597 [hep-th]}}.

\bibitem{Chen:2016hrz}
X.~Chen, Y.~Wang, and Z.-Z. Xianyu, ``{Standard Model Mass Spectrum in
  Inflationary Universe},''
  \href{http://dx.doi.org/10.1007/JHEP04(2017)058}{{\em JHEP} {\bfseries 04}
  (2017) 058},
\href{http://arxiv.org/abs/1612.08122}{{\ttfamily arXiv:1612.08122 [hep-th]}}.

\bibitem{Chua:2018dqh}
W.~Z. Chua, Q.~Ding, Y.~Wang, and S.~Zhou, ``{Imprints of Schwinger Effect on
  Primordial Spectra},'' \href{http://dx.doi.org/10.1007/JHEP04(2019)066}{{\em
  JHEP} {\bfseries 04} (2019) 066},
\href{http://arxiv.org/abs/1810.09815}{{\ttfamily arXiv:1810.09815 [hep-th]}}.

\bibitem{Li:2019ves}
L.~Li, T.~Nakama, C.~M. Sou, Y.~Wang, and S.~Zhou, ``{Gravitational Production
  of Superheavy Dark Matter and Associated Cosmological Signatures},''
  \href{http://dx.doi.org/10.1007/JHEP07(2019)067}{{\em JHEP} {\bfseries 07}
  (2019) 067}, \href{http://arxiv.org/abs/1903.08842}{{\ttfamily
  arXiv:1903.08842 [astro-ph.CO]}}.

\bibitem{Lu:2019tjj}
S.~Lu, Y.~Wang, and Z.-Z. Xianyu, ``{A Cosmological Higgs Collider},''
  \href{http://dx.doi.org/10.1007/JHEP02(2020)011}{{\em JHEP} {\bfseries 02}
  (2020) 011}, \href{http://arxiv.org/abs/1907.07390}{{\ttfamily
  arXiv:1907.07390 [hep-th]}}.

\bibitem{Hook:2019vcn}
A.~Hook, J.~Huang, and D.~Racco, ``{Minimal signatures of the Standard Model in
  non-Gaussianities},''
  \href{http://dx.doi.org/10.1103/PhysRevD.101.023519}{{\em Phys. Rev. D}
  {\bfseries 101} no.~2, (2020) 023519},
  \href{http://arxiv.org/abs/1908.00019}{{\ttfamily arXiv:1908.00019
  [hep-ph]}}.

\bibitem{Wang:2019gbi}
L.-T. Wang and Z.-Z. Xianyu, ``{In Search of Large Signals at the Cosmological
  Collider},'' \href{http://dx.doi.org/10.1007/JHEP02(2020)044}{{\em JHEP}
  {\bfseries 02} (2020) 044},
\href{http://arxiv.org/abs/1910.12876}{{\ttfamily arXiv:1910.12876 [hep-ph]}}.

\bibitem{Li:2020xwr}
L.~Li, S.~Lu, Y.~Wang, and S.~Zhou, ``{Cosmological Signatures of Superheavy
  Dark Matter},'' \href{http://dx.doi.org/10.1007/JHEP07(2020)231}{{\em JHEP}
  {\bfseries 07} (2020) 231}, \href{http://arxiv.org/abs/2002.01131}{{\ttfamily
  arXiv:2002.01131 [hep-ph]}}.

\bibitem{Wang:2020ioa}
L.-T. Wang and Z.-Z. Xianyu, ``{Gauge Boson Signals at the Cosmological
  Collider},'' \href{http://dx.doi.org/10.1007/JHEP11(2020)082}{{\em JHEP}
  {\bfseries 11} (2020) 082}, \href{http://arxiv.org/abs/2004.02887}{{\ttfamily
  arXiv:2004.02887 [hep-ph]}}.

\bibitem{Aoki:2020zbj}
S.~Aoki and M.~Yamaguchi, ``{Disentangling mass spectra of multiple fields in
  cosmological collider},''
  \href{http://dx.doi.org/10.1007/JHEP04(2021)127}{{\em JHEP} {\bfseries 04}
  (2021) 127}, \href{http://arxiv.org/abs/2012.13667}{{\ttfamily
  arXiv:2012.13667 [hep-th]}}.

\bibitem{Cui:2021iie}
Y.~Cui and Z.-Z. Xianyu, ``{Probing Leptogenesis with the Cosmological
  Collider},'' \href{http://dx.doi.org/10.1103/PhysRevLett.129.111301}{{\em
  Phys. Rev. Lett.} {\bfseries 129} no.~11, (2022) 111301},
  \href{http://arxiv.org/abs/2112.10793}{{\ttfamily arXiv:2112.10793
  [hep-ph]}}.

\bibitem{Ema:2023dxm}
Y.~Ema and S.~Verner, ``{Cosmological collider signatures of Higgs-R$^{2}$
  inflation},'' \href{http://dx.doi.org/10.1088/1475-7516/2024/04/039}{{\em
  JCAP} {\bfseries 04} (2024) 039},
  \href{http://arxiv.org/abs/2309.10841}{{\ttfamily arXiv:2309.10841
  [hep-ph]}}.

\bibitem{Bodas:2025vpb}
A.~Bodas, E.~Broadberry, R.~Sundrum, and Z.~Xu, ``{Charged loops at the
  cosmological collider with chemical potential},''
  \href{http://dx.doi.org/10.1007/JHEP01(2026)083}{{\em JHEP} {\bfseries 01}
  (2026) 083}, \href{http://arxiv.org/abs/2507.22978}{{\ttfamily
  arXiv:2507.22978 [hep-ph]}}.

\bibitem{You:2026xoq}
J.~You, L.~Song, C.~Han, H.-J. He, X.~Chen, and Z.-Z. Xianyu, ``{Cosmological
  Collider Signatures from Right-Handed Neutrino Loop},''
  \href{http://arxiv.org/abs/2605.21419}{{\ttfamily arXiv:2605.21419
  [hep-ph]}}.

\bibitem{Chen:2018xck}
X.~Chen, Y.~Wang, and Z.-Z. Xianyu, ``{Neutrino Signatures in Primordial
  Non-Gaussianities},'' \href{http://dx.doi.org/10.1007/JHEP09(2018)022}{{\em
  JHEP} {\bfseries 09} (2018) 022},
\href{http://arxiv.org/abs/1805.02656}{{\ttfamily arXiv:1805.02656 [hep-ph]}}.

\bibitem{Hook:2019zxa}
A.~Hook, J.~Huang, and D.~Racco, ``{Searches for other vacua. Part II. A new
  Higgstory at the cosmological collider},''
  \href{http://dx.doi.org/10.1007/JHEP01(2020)105}{{\em JHEP} {\bfseries 01}
  (2020) 105}, \href{http://arxiv.org/abs/1907.10624}{{\ttfamily
  arXiv:1907.10624 [hep-ph]}}.

\bibitem{Tong:2023krn}
X.~Tong, Y.~Wang, C.~Zhang, and Y.~Zhu, ``{BCS in the sky: signatures of
  inflationary fermion condensation},''
  \href{http://dx.doi.org/10.1088/1475-7516/2024/04/022}{{\em JCAP} {\bfseries
  04} (2024) 022}, \href{http://arxiv.org/abs/2304.09428}{{\ttfamily
  arXiv:2304.09428 [hep-th]}}.

\bibitem{Aoki:2026olh}
S.~Aoki and A.~Strumia, ``{Cosmological collider signals of modular spontaneous
  CP breaking},'' \href{http://arxiv.org/abs/2604.05548}{{\ttfamily
  arXiv:2604.05548 [hep-ph]}}.

\bibitem{You:2024hit}
J.~You, L.~Song, H.-J. He, and C.~Han, ``{Cosmological non-Gaussianity from the
  neutrino seesaw mechanism},'' \href{http://dx.doi.org/10.1103/lhmk-ycfr}{{\em
  Phys. Rev. D} {\bfseries 112} no.~8, (2025) 083555},
  \href{http://arxiv.org/abs/2412.16033}{{\ttfamily arXiv:2412.16033
  [hep-ph]}}.

\bibitem{Han:2024qbw}
C.~Han, H.-J. He, L.~Song, and J.~You, ``{Cosmological signatures of neutrino
  seesaw mechanism},'' \href{http://dx.doi.org/10.1103/b7hv-3h2p}{{\em Phys.
  Rev. D} {\bfseries 112} no.~8, (2025) L081309},
  \href{http://arxiv.org/abs/2412.21045}{{\ttfamily arXiv:2412.21045
  [hep-ph]}}.

\bibitem{Sou:2021juh}
C.~M. Sou, X.~Tong, and Y.~Wang, ``{Chemical-potential-assisted particle
  production in FRW spacetimes},''
  \href{http://dx.doi.org/10.1007/JHEP06(2021)129}{{\em JHEP} {\bfseries 06}
  (2021) 129}, \href{http://arxiv.org/abs/2104.08772}{{\ttfamily
  arXiv:2104.08772 [hep-th]}}.

\bibitem{Aoki:2026gge}
S.~Aoki and M.~Yamaguchi, ``{Hubble Susceptibility: Beyond Effective-Mass
  Spectroscopy in the Cosmological Collider},''
  \href{http://arxiv.org/abs/2610.26014}{{\ttfamily arXiv:2610.26014
  [hep-th]}}.

\bibitem{Marolf:2010zp}
D.~Marolf and I.~A. Morrison, ``{The IR stability of de Sitter: Loop
  corrections to scalar propagators},''
  \href{http://dx.doi.org/10.1103/PhysRevD.82.105032}{{\em Phys. Rev. D}
  {\bfseries 82} (2010) 105032},
  \href{http://arxiv.org/abs/1006.0035}{{\ttfamily arXiv:1006.0035 [gr-qc]}}.

\bibitem{DiPietro:2021sjt}
L.~Di~Pietro, V.~Gorbenko, and S.~Komatsu, ``{Analyticity and unitarity for
  cosmological correlators},''
  \href{http://dx.doi.org/10.1007/JHEP03(2022)023}{{\em JHEP} {\bfseries 03}
  (2022) 023}, \href{http://arxiv.org/abs/2108.01695}{{\ttfamily
  arXiv:2108.01695 [hep-th]}}.

\bibitem{Sleight:2021plv}
C.~Sleight and M.~Taronna, ``{From dS to AdS and back},''
  \href{http://dx.doi.org/10.1007/JHEP12(2021)074}{{\em JHEP} {\bfseries 12}
  (2021) 074}, \href{http://arxiv.org/abs/2109.02725}{{\ttfamily
  arXiv:2109.02725 [hep-th]}}.

\bibitem{Loparco:2023rug}
M.~Loparco, J.~Penedones, K.~Salehi~Vaziri, and Z.~Sun, ``{The
  K\"all\'en-Lehmann representation in de Sitter spacetime},''
  \href{http://dx.doi.org/10.1007/JHEP12(2023)159}{{\em JHEP} {\bfseries 12}
  (2023) 159}, \href{http://arxiv.org/abs/2306.00090}{{\ttfamily
  arXiv:2306.00090 [hep-th]}}.

\bibitem{Tong:2021wai}
X.~Tong, Y.~Wang, and Y.~Zhu, ``{Cutting rule for cosmological collider
  signals: a bulk evolution perspective},''
  \href{http://dx.doi.org/10.1007/JHEP03(2022)181}{{\em JHEP} {\bfseries 03}
  (2022) 181}, \href{http://arxiv.org/abs/2112.03448}{{\ttfamily
  arXiv:2112.03448 [hep-th]}}.

\bibitem{Ema:2024hkj}
Y.~Ema and K.~Mukaida, ``{Cutting rule for in-in correlators and cosmological
  collider},'' \href{http://dx.doi.org/10.1007/JHEP12(2024)194}{{\em JHEP}
  {\bfseries 12} (2024) 194}, \href{http://arxiv.org/abs/2409.07521}{{\ttfamily
  arXiv:2409.07521 [hep-th]}}.

\end{thebibliography}\endgroup

\clearpage
\onecolumngrid
\appendix
\setcounter{secnumdepth}{2}

\section{Proof of the exclusion rule for one-loop \texorpdfstring{$N$}{N}-gons}
\label{sec:ngon}
Throughout the appendices, we set $H=1$.
For a scalar of mass parameter $\nu = \sqrt{m^2-9/4}$, the normalised Bunch-Davies (BD) mode function can be chosen as
\begin{align}
  u(k,\tau) =\frac{\sqrt{\pi}}{2}e^{-\pi\nu/2}(-\tau)^{3/2} \mathrm{H}^{(1)}_{\ii\nu}(-k\tau) =\sum_{\mathsf{c}=\pm1}k^{\ii\mathsf{c}\nu}f_{\mathsf{c}}(k,\tau)\,, \label{Hankel}
\end{align}
where the functions $f_\pm$ are
\begin{align}
  f_{\mathsf{c}}(k,\tau) \equiv\frac{\mathsf{c}\sqrt{\pi}e^{\mathsf{c}\pi\nu/2}} {2^{1+\ii\mathsf{c}\nu}\sinh(\pi\nu)} (-\tau)^{3/2+\ii\mathsf{c}\nu} \sum_{n=0}^{\infty} \frac{(-k^2\tau^2/4)^n} {n!\,\Gamma(n+1+\ii\mathsf{c}\nu)}\,, \label{frobenius}
\end{align}
which admit Taylor expansions in $k$ around the origin and obey $f_{\mathsf{c}}^*=-e^{\mathsf{c}\pi\nu}f_{-\mathsf{c}}$. The four SK bulk-to-bulk propagators are constructed from this mode function as follows:
\begin{align}
  D_{-+}(k;\tau_1,\tau_2) &=u(k,\tau_1)u^*(k,\tau_2)\,, \label{Dmp}\\
  D_{+-}(k;\tau_1,\tau_2) &=u^*(k,\tau_1)u(k,\tau_2)\,,\\
  D_{++}(k;\tau_1,\tau_2) &=\theta(\tau_1-\tau_2)D_{-+}(k;\tau_1,\tau_2) +\theta(\tau_2-\tau_1)D_{+-}(k;\tau_1,\tau_2)\,, \\
  D_{--}(k;\tau_1,\tau_2) &=\theta(\tau_2-\tau_1)D_{-+}(k;\tau_1,\tau_2) +\theta(\tau_1-\tau_2)D_{+-}(k;\tau_1,\tau_2)\,. \label{Dmm}
\end{align}

For later convenience, we reproduce the integral expression for the $N$-gon correlator in \eqref{TN} and \eqref{QN}:
\begin{align}
  \mathcal T_N &\sim \sum_{\mathsf a_1,\cdots,\mathsf a_N=\pm} \left[\prod_{v=1}^{N-1}\int_{-\infty}^{0} \ii\mathsf a_v\dd\tau_v\,(-\tau_v)^{p_v} e^{\ii\mathsf a_vE_v\tau_v}\right] \int_{-\infty}^{0}\ii\mathsf a_N\dd\tau_N\, (-\tau_N)^{p_N}e^{\ii\mathsf a_Nk_L\tau_N}\,\mathcal Q\,, \label{TNfull}
\end{align}
where
\begin{align}
  \mathcal Q&\equiv \int\frac{\dd^3\bm q}{(2\pi)^3}\, D^{(\nu_1)}_{\mathsf a_1\mathsf a_N}(q;\tau_1,\tau_N) D^{(\nu_N)}_{\mathsf a_N\mathsf a_{N-1}}(|\bm q+\bm k_L|;\tau_N,\tau_{N-1})\times \prod_{i=2}^{N-1} D^{(\nu_i)}_{\mathsf a_{i-1}\mathsf a_i} (|\bm q-\sum_{\ell=1}^{i-1}\bm k_\ell|; \tau_{i-1},\tau_i)\,. \label{QNfull}
\end{align}
Here $E_v$ and $\bm k_v$ are the total external energy and momentum entering vertex $v$, and $p_v$ encodes the schematic vertex time dependence. Overall couplings and fixed external normalisations are suppressed.
We evaluate the loop integrals in dimensional regularisation, taking the physical limit $d\to3$ after subtracting any UV divergence.
Scaleless integrals vanish in this prescription.

We take the generic single soft limit $k_L/k_S\to0$ defined in the main text, with $k_S>0$ fixed independently of $k_L$. The hard momenta approach a fixed configuration analytically while preserving momentum conservation, and no additional independent momentum channel becomes soft. More explicitly, both the hard vertex energies and the sums of external momenta entering consecutive vertices obey
\begin{align}
  E_v&\gtrsim k_S\,, \qquad v=1,\cdots,N-1\,,\\
  |\sum_{\ell=i}^{j}\bm k_\ell|&\gtrsim k_S\,, \qquad 1\le i\le j\le N-1,\quad (i,j)\ne(1,N-1)\,. \label{kinematics}
\end{align}
In particular, $i=j$ gives $k_v\gtrsim k_S$ at each hard vertex with $v\neq N$. The excluded sum satisfies $\sum_{\ell=1}^{N-1}\bm k_\ell=-\bm k_L$ and therefore represents the chosen soft channel.

\subsection{Validity of the directional cut}
\label{sec:cutting}

CC signals can be identified as the discontinuity associated with the soft limit and can be extracted using appropriate cutting rules. For the single soft signal considered here, we apply a directional cut to the two propagators attached to the soft vertex. This cut was introduced for triangle diagrams in \cite{Qin:2026yuo}, and we now extend it to a general $N$-gon.
Writing $\bm p\equiv\bm q+\bm k_L$ and $p\equiv|\bm p|$, each of the two propagators attached to $\tau_N$ in \eqref{QNfull} splits into two parts:
  \begin{align}
  D^{(\nu_1)}_{\mathsf a_1\mathsf a_N}(q;\tau_1,\tau_N) &= D^{(\nu_1)}_{-\mathsf a_N,\mathsf a_N}(q;\tau_1,\tau_N) \nonumber\\
  &+\delta_{\mathsf a_1\mathsf a_N}\theta(\tau_N-\tau_1) \left[D^{(\nu_1)}_{\mathsf a_N,-\mathsf a_N}(q;\tau_1,\tau_N) -D^{(\nu_1)}_{-\mathsf a_N,\mathsf a_N}(q;\tau_1,\tau_N)\right]\,, \label{cutq} \\
  D^{(\nu_N)}_{\mathsf a_N\mathsf a_{N-1}}(p;\tau_N,\tau_{N-1}) &= D^{(\nu_N)}_{\mathsf a_N,-\mathsf a_N}(p;\tau_N,\tau_{N-1}) \nonumber\\
  &+\delta_{\mathsf a_N\mathsf a_{N-1}}\theta(\tau_N-\tau_{N-1}) \left[D^{(\nu_N)}_{-\mathsf a_N,\mathsf a_N}(p;\tau_N,\tau_{N-1}) -D^{(\nu_N)}_{\mathsf a_N,-\mathsf a_N}(p;\tau_N,\tau_{N-1})\right]\,. \label{cutp}
\end{align}
The first term in each of \eqref{cutq} and \eqref{cutp} is the \emph{cut part}, while the second is the \emph{time-ordered part}.
The product of the two cut parts defines the \emph{cut contribution} to \eqref{QNfull}.
Each of the other three \emph{time-ordered terms} contains at least one step function multiplying a commutator $D_{+-}-D_{-+}$. Below we show that these time-ordered terms admit Taylor expansions about $k_L=0$ and thus have no CC signal.

In particular, consider the time-ordered contribution $\mathcal T_{N,\mathrm{TO}}$ obtained by replacing $D^{(\nu_N)}_{\mathsf a_N\mathsf a_{N-1}}$ with its time-ordered part, namely the second term on the right-hand side of \eqref{cutp}, while keeping the other propagators unchanged. In the rest of this subsection, $\mathcal Q_{\mathrm{TO}}$ and $\mathcal P_{\mathrm{TO}}$ denote the loop integral and integrand of this contribution:
\begin{align}
  \mathcal T_{N,\mathrm{TO}}= \sum_{\mathsf a_1,\cdots,\mathsf a_N=\pm} \delta_{\mathsf a_N\mathsf a_{N-1}} \left[\prod_{v=1}^{N-1}\int_{-\infty}^{0} \ii\mathsf a_v\dd\tau_v\,(-\tau_v)^{p_v} e^{\ii\mathsf a_vE_v\tau_v}\right] \int_{\tau_{N-1}}^{0}\ii\mathsf a_N\dd\tau_N\, (-\tau_N)^{p_N}e^{\ii\mathsf a_Nk_L\tau_N} \,\mathcal Q_{\mathrm{TO}}\,, \label{TTO}
\end{align}
where the step function has been absorbed into the lower limit, so the $\tau_N$ interval is finite for fixed hard vertex times. The loop integral is
\begin{align}
  \mathcal Q_{\mathrm{TO}}&\equiv \int\frac{\dd^3\bm q}{(2\pi)^3}\, \mathcal P_{\mathrm{TO}}(\bm q,\bm k_L)\,, \label{QTO}
\end{align}
with
\begin{align}
  \mathcal P_{\mathrm{TO}}(\bm q,\bm k_L)&\equiv D^{(\nu_1)}_{\mathsf a_1\mathsf a_N}(q;\tau_1,\tau_N) \times\left[D^{(\nu_N)}_{-\mathsf a_N,\mathsf a_N}(p;\tau_N,\tau_{N-1}) -D^{(\nu_N)}_{\mathsf a_N,-\mathsf a_N}(p;\tau_N,\tau_{N-1})\right]\nonumber\\
  &\times\prod_{i=2}^{N-1} D^{(\nu_i)}_{\mathsf a_{i-1}\mathsf a_i} (|\bm q-\sum_{m=1}^{i-1}\bm k_m|; \tau_{i-1},\tau_i)\,. \label{PTO}
\end{align}
At fixed hard vertex times, the factor $e^{\ii\mathsf a_Nk_L\tau_N}$ in \eqref{TTO} admits a Taylor expansion in $-k_L\tau_N$ uniformly over the finite $\tau_N$ interval and hence supplies only nonnegative integer powers of $k_L$. The remaining hard time integrations are performed term by term in this series, first in a common convergence domain and then by analytic continuation, preserving this integer power structure. Any nonanalytic dependence on $k_L$ in this limit must therefore arise from the loop integral $\mathcal Q_{\mathrm{TO}}$. We next analyse the analytic properties of $\mathcal Q_{\mathrm{TO}}$ in the soft limit $k_L\to0$ at fixed vertex times.

Following the method of regions, the asymptotic expansion as $k_L/k_S\to0$ takes the form $\mathcal Q_{\mathrm{TO}}\sim\mathcal Q_{\mathrm{TO},1}+\mathcal Q_{\mathrm{TO},2}+\mathcal Q_{\mathrm{TO},3}$, where the three contributions correspond to the three momentum regions described below. Each contribution is defined by expanding $\mathcal P_{\mathrm{TO}}$ in the corresponding region and integrating term by term over all loop momentum space in dimensional regularisation.

\emph{(i) All internal lines hard.} As all internal momentum magnitudes remain nonzero in this region, $\mathcal P_{\mathrm{TO}}$ admits an ordinary Taylor expansion in $\bm k_L$:
\begin{align}
  \mathcal P_{\mathrm{TO}}(\bm q,\bm k_L) \sim\sum_{n=0}^{\infty}\frac{1}{n!} \left[(\bm k_L\cdot\nabla_{\bm k})^n \mathcal P_{\mathrm{TO}}(\bm q,\bm k)\right]_{\bm k=\bm0}\,, \label{Phard}
\end{align}
so the integral over $\bm q$ can also be expanded in $\bm k_L$ as a Taylor series
\begin{align}
  \mathcal Q_{\mathrm{TO},1} \sim\sum_{n=0}^{\infty}\frac{1}{n!} \int\frac{\dd^3\bm q}{(2\pi)^3}\, \left[(\bm k_L\cdot\nabla_{\bm k})^n \mathcal P_{\mathrm{TO}}(\bm q,\bm k)\right]_{\bm k=\bm0}\,,
\end{align}
where every order contains only integer powers of the soft momentum $\bm k_L$.

\emph{(ii) One internal line soft.} Since $\bm p-\bm q=\bm k_L$, soft $q$ necessarily implies soft $p$, and vice versa, so neither can be the isolated soft line. Let us then take the $j$-th line with $2\le j\le N-1$ to be soft and shift the loop momentum to
\begin{align}
  \bm\ell=\bm q-\sum_{i=1}^{j-1}\bm k_i,\qquad \ell=|\bm\ell|\,,
\end{align}
so the loop integrand becomes
\begin{align}
  \mathcal P_{\mathrm{TO}}(\bm\ell+\sum_{m=1}^{j-1}\bm k_m, \bm k_L) &= D^{(\nu_j)}_{\mathsf a_{j-1}\mathsf a_j}(\ell;\tau_{j-1},\tau_j) D^{(\nu_1)}_{\mathsf a_1\mathsf a_N} (|\bm\ell+\sum_{m=1}^{j-1}\bm k_m|; \tau_1,\tau_N) \nonumber\\
  &\times\left[ D^{(\nu_N)}_{-\mathsf a_N,\mathsf a_N} (|\bm\ell+\bm k_L+\sum_{m=1}^{j-1}\bm k_m|;\tau_N,\tau_{N-1}) -D^{(\nu_N)}_{\mathsf a_N,-\mathsf a_N} (|\bm\ell+\bm k_L+\sum_{m=1}^{j-1}\bm k_m|;\tau_N,\tau_{N-1}) \right] \nonumber\\
  &\times\prod_{\substack{i=2\\i\ne j}}^{N-1} D^{(\nu_i)}_{\mathsf a_{i-1}\mathsf a_i} (|\bm\ell+\sum_{m=1}^{j-1}\bm k_m -\sum_{m=1}^{i-1}\bm k_m|;\tau_{i-1},\tau_i)\,.
\end{align}
All factors except the $j$-th propagator remain hard and admit Taylor expansions in $\bm\ell$, as in \eqref{Phard}, while the soft $j$-th propagator has a Frobenius expansion, for example,
\begin{align}
  D^{(\nu_j)}_{-+}(\ell;\tau_{j-1},\tau_j) &= -\sum_{\mathsf{c}=\pm1}e^{-\mathsf{c}\pi\nu_j}\ell^{2\ii\mathsf{c}\nu_j} f_{\mathsf{c}}^{(\nu_j)}(\ell,\tau_{j-1}) f_{\mathsf{c}}^{(\nu_j)}(\ell,\tau_j)\nonumber\\
  &~~~-\sum_{\mathsf{c}=\pm1}e^{\mathsf{c}\pi\nu_j} f_{\mathsf{c}}^{(\nu_j)}(\ell,\tau_{j-1}) f_{-\mathsf{c}}^{(\nu_j)}(\ell,\tau_j)\,. \label{Dsoft}
\end{align}
Here $f_{\mathsf{c}}^{(\nu_j)}$ is given by \eqref{frobenius} with $\nu=\nu_j$, and the superscript labels the mass parameter. All four SK propagators contain the same powers $1,\ell^{\pm2\ii\nu_j}$ with coefficients analytic in $\ell^2$. Combining this Frobenius expansion with the Taylor expansions of the hard factors gives
\begin{align}
  \mathcal P_{\mathrm{TO}}(\bm\ell+\sum_{m=1}^{j-1}\bm k_m, \bm k_L) \sim\sum_{\alpha\in\{0,\pm2\ii\nu_j\}} \sum_{n=0}^\infty \ell^\alpha P_{j,\alpha}^{(n)}(\bm\ell,\bm k_L)\,,
\end{align}
where $P_{j,\alpha}^{(n)}$ is a homogeneous polynomial of degree $n$ in $\bm\ell$. Each term therefore gives a scaleless integral over all loop momentum space, which vanishes in dimensional regularisation. Thus this region contributes zero order by order:
\begin{align}
  \mathcal Q_{\mathrm{TO},2} = \sum_{j=2}^{N-1} \int \frac{\dd^3\bm\ell}{(2\pi)^3}\, \sum_{\alpha\in\{0,\pm2\ii\nu_j\}} \sum_{n=0}^\infty \ell^\alpha P_{j,\alpha}^{(n)}(\bm\ell,\bm k_L) = 0\,.
\end{align}

\emph{(iii) Two internal lines soft.} The only allowed pair is $q,p=\mathcal O(k_L)$, any other pair, or three or more soft lines, would violate \eqref{kinematics}.
Using $\bm p$ as the loop variable gives
\begin{align}
  \mathcal P_{\mathrm{TO}}(\bm p-\bm k_L,\bm k_L)&= D^{(\nu_1)}_{\mathsf a_1\mathsf a_N} (|\bm p-\bm k_L|;\tau_1,\tau_N) \left[D^{(\nu_N)}_{-\mathsf a_N,\mathsf a_N}(p;\tau_N,\tau_{N-1}) -D^{(\nu_N)}_{\mathsf a_N,-\mathsf a_N}(p;\tau_N,\tau_{N-1})\right] \nonumber\\
  &\times\prod_{i=2}^{N-1} D^{(\nu_i)}_{\mathsf a_{i-1}\mathsf a_i} (|\bm p-\bm k_L-\sum_{m=1}^{i-1}\bm k_m|; \tau_{i-1},\tau_i)\,.
\end{align}
Using \eqref{Hankel}, we obtain
\begin{align}
  &D^{(\nu_N)}_{-\mathsf a_N,\mathsf a_N}(p;\tau_N,\tau_{N-1}) -D^{(\nu_N)}_{\mathsf a_N,-\mathsf a_N}(p;\tau_N,\tau_{N-1}) \nonumber\\
  &\qquad=-2\mathsf a_N\sinh(\pi\nu_N) \bigl[f_{+}^{(\nu_N)}(p,\tau_N)f_{-}^{(\nu_N)}(p,\tau_{N-1}) -f_{-}^{(\nu_N)}(p,\tau_N)f_{+}^{(\nu_N)}(p,\tau_{N-1})\bigr]\,,
\end{align}
which admits a Taylor expansion in $p^2$ at the soft limit. As in \eqref{Dsoft}, the $q$-line propagator contains the powers $q^\alpha=|\bm p-\bm k_L|^\alpha$, with $\alpha\in\{0,\pm2\ii\nu_1\}$, multiplied by coefficients analytic in $q^2$. After factoring out these powers, all remaining factors admit joint Taylor expansions in both $\bm p$ and $\bm k_L$ at fixed times. Combining these expansions gives
\begin{align}
  \mathcal P_{\mathrm{TO}}(\bm p-\bm k_L,\bm k_L) \sim\sum_{\alpha\in\{0,\pm2\ii\nu_1\}}\sum_{n=0}^{\infty} |\bm p-\bm k_L|^\alpha Q_\alpha^{(n)}(\bm p,\bm k_L)\,,
\end{align}
where $Q_\alpha^{(n)}$ is a homogeneous polynomial of total degree $n$ in $\bm p$ and $\bm k_L$. As in region (ii), this region contributes zero order by order:
\begin{align}
  \mathcal Q_{\mathrm{TO},3} &=\sum_{\alpha\in\{0,\pm2\ii\nu_1\}}\sum_{n=0}^{\infty} \int\frac{\dd^3\bm p}{(2\pi)^3}\, Q_\alpha^{(n)}(\bm p,\bm k_L)|\bm p-\bm k_L|^\alpha \nonumber\\
  &=\sum_{\alpha\in\{0,\pm2\ii\nu_1\}}\sum_{n=0}^{\infty} \int\frac{\dd^3\bm q}{(2\pi)^3}\, \widetilde Q_\alpha^{(n)}(\bm q,\bm k_L)q^\alpha=0\,.
\end{align}
In the second line we shifted $\bm q=\bm p-\bm k_L$ and defined $\widetilde Q_\alpha^{(n)}(\bm q,\bm k_L)\equiv Q_\alpha^{(n)}(\bm q+\bm k_L,\bm k_L)$. Since $\widetilde Q_\alpha^{(n)}$ remains polynomial, each monomial multiplied by $q^\alpha$ gives a scaleless integral over $\bm q$, which vanishes in dimensional regularisation. Therefore, $\mathcal Q_{\mathrm{TO}}\sim\mathcal Q_{\mathrm{TO},1}+\mathcal Q_{\mathrm{TO},2}+\mathcal Q_{\mathrm{TO},3}$ contains only terms analytic in $\bm k_L$ at fixed vertex times.

Thus $\mathcal T_{N,\mathrm{TO}}$ carries no CC signal. The same argument applies to the other time-ordered terms, establishing the validity of the directional cut.

\subsection{Matching to auxiliary bubbles}

We now focus on the CC signal and apply the directional cut to the integrand in \eqref{QNfull}:
\begin{align}
  \mathcal Q_{\mathrm{cut}} &\equiv\int\frac{\dd^3\bm q}{(2\pi)^3}\,\mathcal P_{\mathrm{cut}}\,, \nonumber\\
  \mathcal P_{\mathrm{cut}} &=D^{(\nu_1)}_{-\mathsf a_N,\mathsf a_N} (q;\tau_1,\tau_N) D^{(\nu_N)}_{\mathsf a_N,-\mathsf a_N} (p;\tau_N,\tau_{N-1}) \prod_{i=2}^{N-1} D^{(\nu_i)}_{\mathsf a_{i-1}\mathsf a_i} (|\bm q-\sum_{\ell=1}^{i-1}\bm k_\ell|; \tau_{i-1},\tau_i)\,.
\end{align}
The analysis of Sec.~\ref{sec:cutting} also applies to the cut contribution: the all hard region is analytic in the soft momentum, while an isolated soft line gives scaleless integrals order by order. Thus only the region $p,q=\mathcal O(k_L)$ can contribute a CC signal.

We expand the dependence on the soft internal momenta at fixed hard kinematic parameters, retaining the hard external phases $e^{\ii\mathsf a_vE_v\tau_v}$ in the time integrals. The hard propagators are Taylor expanded in their soft momentum shifts, with their full dependence on the hard momenta and vertex times retained.

For definiteness, take $\mathsf a_N=+$ and write $u_{\nu_j}$ for the mode with mass parameter $\nu_j$. Using \eqref{Hankel} and \eqref{frobenius}, we expand the two modes at $\tau_1$ and $\tau_{N-1}$ while keeping both modes at $\tau_N$ exact. The cut loop integrand then has the double expansion
\begin{align}
  \left.\mathcal P_{\mathrm{cut}}\right|_{\mathsf a_N=+} &\sim\sum_{\mathsf{c}_1,\mathsf{c}_2=\pm} \sum_{m,n=0}^{\infty} P_{\mathsf{c}_1\mathsf{c}_2}^{(m,n)}(\bm q,\bm p;\tau_1,\cdots,\tau_{N-1}) \times q^{\ii\mathsf{c}_1\nu_1}p^{\ii\mathsf{c}_2\nu_N} \times u_{\nu_1}^*(q,\tau_N)u_{\nu_N}^*(p,\tau_N)\,. \label{Pcut}
\end{align}
Here $P_{\mathsf{c}_1\mathsf{c}_2}^{(m,n)}$ is homogeneous of degree $m$ in $\bm q$ and degree $n$ in $\bm p$. Suppressing Cartesian indices on its coefficient tensors, we write
\begin{align}
  P_{\mathsf{c}_1\mathsf{c}_2}^{(m,n)}(\bm q,\bm p;\tau_1,\cdots,\tau_{N-1}) \equiv c^{(m,n)}_{\mathsf{c}_1\mathsf{c}_2}(\tau_1,\cdots,\tau_{N-1}) q_{i_1}\cdots q_{i_m}p_{j_1}\cdots p_{j_n}\,. \label{c}
\end{align}
The contraction of the suppressed indices of $c^{(m,n)}$ with the displayed momentum tensors is understood. The coefficients are symmetric within each group of $m$ and $n$ indices. Their hard time dependence is explicit, while their dependence on the hard kinematics and SK indices is suppressed.
Using $f_{\mathsf{c}}^{(\nu)*}=-e^{\mathsf{c}\pi\nu}f_{-\mathsf{c}}^{(\nu)}$ and relabelling the branches, the $\mathsf a_N=-$ expansion has the same coefficients $c^{(m,n)}$, with the soft mode pair replaced by $e^{-\pi(\mathsf{c}_1\nu_1+\mathsf{c}_2\nu_N)}u_{\nu_1}(q,\tau_N)u_{\nu_N}(p,\tau_N)$.

Denoting the cut contribution from the double soft region by $\mathcal T_{N,\mathrm{cut}}$, we substitute \eqref{Pcut} and \eqref{c} into \eqref{TNfull} and sum over $\mathsf a_N=\pm$ to obtain
\begin{align}
  \mathcal T_{N,\mathrm{cut}} &= \sum_{\mathsf a_1,\cdots,\mathsf a_{N-1}=\pm}\sum_{\mathsf{c}_1,\mathsf{c}_2=\pm} \sum_{m,n=0}^{\infty} \left[\prod_{v=1}^{N-1}\int_{-\infty}^{0} \ii\mathsf a_v\dd\tau_v\,(-\tau_v)^{p_v} e^{\ii\mathsf a_vE_v\tau_v}\right] c^{(m,n)}_{\mathsf{c}_1\mathsf{c}_2}(\tau_1,\cdots,\tau_{N-1}) \nonumber\\
  &\times\Biggl[\ii\int_{-\infty}^{0}\dd\tau_N\,(-\tau_N)^{p_N} \int\frac{\dd^3\bm q}{(2\pi)^3}\, q_{i_1}\cdots q_{i_m}p_{j_1}\cdots p_{j_n} q^{\ii\mathsf{c}_1\nu_1}p^{\ii\mathsf{c}_2\nu_N} \nonumber\\
  &\qquad\times\Bigl( e^{\ii k_L\tau_N}u_{\nu_1}^*(q,\tau_N)u_{\nu_N}^*(p,\tau_N) -e^{-\pi(\mathsf{c}_1\nu_1+\mathsf{c}_2\nu_N)}e^{-\ii k_L\tau_N} u_{\nu_1}(q,\tau_N)u_{\nu_N}(p,\tau_N) \Bigr)\Biggr]\,. \label{factorisation}
\end{align}
Rescaling $\bm q\to k_L\bm q$ and $\tau_N\to\tau_N/k_L$ shows that the loop integral together with the $\tau_N$ integral in square brackets scales as $k_L^{m+n-p_N-1+\ii(\mathsf{c}_1\nu_1+\mathsf{c}_2\nu_N)}$. Thus $(\mathsf{c}_1,\mathsf{c}_2)$ fixes the oscillation frequency to be $|\mathsf{c}_1\nu_1+\mathsf{c}_2\nu_N|$, while $m+n$ counts the additional integer powers of $k_L$.

To match \eqref{factorisation}, consider the auxiliary bubble defined, up to overall normalisation and symmetry factors, by
\begin{align}
  \mathcal T_2^{(r,s)} &\equiv \sum_{\mathsf a_1,\mathsf a_N=\pm} \int_{-\infty}^{0}\ii\mathsf a_1\dd\tau_1\, (-\tau_1)^{r+s-2}e^{\ii\mathsf a_1E\tau_1} \int_{-\infty}^{0}\ii\mathsf a_N\dd\tau_N\, (-\tau_N)^{p_N}e^{\ii\mathsf a_Nk_L\tau_N} \nonumber\\
  &\times\int\frac{\dd^3\bm q}{(2\pi)^3}\, q_{\alpha_1}\cdots q_{\alpha_r}p_{\beta_1}\cdots p_{\beta_s} \times D^{(\nu_1)}_{\mathsf a_1\mathsf a_N}(q;\tau_1,\tau_N) D^{(\nu_N)}_{\mathsf a_N\mathsf a_1}(p;\tau_N,\tau_1)\,. \label{T2}
\end{align}
Here the spatial indices $\alpha_1,\cdots,\alpha_r,\beta_1,\cdots,\beta_s$ of $\mathcal T_2^{(r,s)}$ are suppressed. The auxiliary bubble is obtained by replacing the original hard subgraph and all external legs attached to it by a single local (but noncovariant) vertex $\sqrt{-g}\phi^2\mathcal O^{(r,s)}$. The two heavy propagators and the original soft vertex are retained. Here $\phi$ is an auxiliary conformally coupled scalar whose two hard external legs carry momenta $\bm K_1$ and $\bm K_2$ satisfying
\begin{equation}
  K_1+K_2=E\gg k_L,\qquad \bm K_1+\bm K_2=-\bm k_L\,,
\end{equation}
and the local noncovariant heavy field bilinear, with its spatial indices suppressed, is
\begin{align}
  \mathcal O^{(r,s)} =\frac{(-\ii)^r\ii^s}{a^{r+s}(\tau_1)} (\partial_{\alpha_1}\cdots\partial_{\alpha_r}\sigma_1) (\partial_{\beta_1}\cdots\partial_{\beta_s}\sigma_N)\,. \label{Ors}
\end{align}

Similarly, the CC signal of $\mathcal T_2^{(r,s)}$ can be extracted using the directional cutting rule of Sec.~\ref{sec:cutting}. Denoting its cut contribution from the double soft region by $\mathcal T_{2,\mathrm{cut}}^{(r,s)}$, we obtain
\begin{align}
  \mathcal T_{2,\mathrm{cut}}^{(r,s)} &= \sum_{\mathsf a_1=\pm}\sum_{m,n=0}^{\infty}\sum_{\mathsf{c}_1,\mathsf{c}_2=\pm} \int_{-\infty}^{0}\ii\mathsf a_1\dd\tau_1\, (-\tau_1)^{r+s-2}e^{\ii\mathsf a_1E\tau_1} d^{(r,s;m,n)}_{\mathsf{c}_1\mathsf{c}_2}(\tau_1) \nonumber\\
  &\times\Biggl[\ii\int_{-\infty}^{0}\dd\tau_N\,(-\tau_N)^{p_N} \int\frac{\dd^3\bm q}{(2\pi)^3}\, q_{i_1}\cdots q_{i_m}p_{j_1}\cdots p_{j_n} q^{\ii\mathsf{c}_1\nu_1}p^{\ii\mathsf{c}_2\nu_N} \nonumber\\
  &\qquad\times\Bigl( e^{\ii k_L\tau_N}u_{\nu_1}^*(q,\tau_N)u_{\nu_N}^*(p,\tau_N) -e^{-\pi(\mathsf{c}_1\nu_1+\mathsf{c}_2\nu_N)}e^{-\ii k_L\tau_N} u_{\nu_1}(q,\tau_N)u_{\nu_N}(p,\tau_N) \Bigr)\Biggr]\,, \label{T2cut}
\end{align}
where, for fixed $r,s$, the coefficient tensors $d^{(r,s;m,n)}$ are defined by
\begin{align}
  &q_{\alpha_1}\cdots q_{\alpha_r}p_{\beta_1}\cdots p_{\beta_s} f_{\mathsf{c}_1}^{(\nu_1)}(q,\tau_1) f_{\mathsf{c}_2}^{(\nu_N)}(p,\tau_1) \equiv \sum_{m,n=0}^{\infty} d^{(r,s;m,n)}_{\mathsf{c}_1\mathsf{c}_2}(\tau_1) q_{i_1}\cdots q_{i_m}p_{j_1}\cdots p_{j_n}\,. \label{d}
\end{align}
For each nonzero frequency branch $(\mathsf{c}_1,\mathsf{c}_2)$, we determine $g^{(r,s)}_{\mathsf{c}_1\mathsf{c}_2}$ by matching the hard time coefficients in \eqref{factorisation} and \eqref{T2cut} order by order in $k_L$, before evaluating the common loop and $\tau_N$ integrals.
In particular, matching the leading hard time coefficients fixes $g^{(0,0)}_{\mathsf{c}_1\mathsf{c}_2}$ through
\begin{align}
  &g^{(0,0)}_{\mathsf{c}_1\mathsf{c}_2} \sum_{\mathsf a_1=\pm}\ii\mathsf a_1\int_{-\infty}^{0}\dd\tau_1\, (-\tau_1)^{-2}e^{\ii\mathsf a_1E\tau_1} d^{(0,0;0,0)}_{\mathsf{c}_1\mathsf{c}_2}(\tau_1) \nonumber\\
  &\quad=\sum_{\mathsf a_1,\cdots,\mathsf a_{N-1}=\pm} \left[\prod_{v=1}^{N-1}\int_{-\infty}^{0} \ii\mathsf a_v\dd\tau_v\,(-\tau_v)^{p_v} e^{\ii\mathsf a_vE_v\tau_v}\right] c^{(0,0)}_{\mathsf{c}_1\mathsf{c}_2}(\tau_1,\cdots,\tau_{N-1})\,. \label{matching0}
\end{align}

\eqref{matching0} matches the leading term in $\mathcal T_{N,\mathrm{cut}}$ carrying the oscillatory factor $(k_L/k_S)^{\ii(\mathsf{c}_1\nu_1+\mathsf{c}_2\nu_N)}$ to the corresponding term in $g^{(0,0)}_{\mathsf{c}_1\mathsf{c}_2}\mathcal T_{2,\mathrm{cut}}^{(0,0)}$, generated by the hard vertex $\sqrt{-g}\,g^{(0,0)}_{\mathsf{c}_1\mathsf{c}_2}\phi^2\mathcal O^{(0,0)}$. This auxiliary bubble satisfies case (i) of the exclusion rule, so its CC signal at this frequency vanishes. The matched term in $\mathcal T_{N,\mathrm{cut}}$ therefore also vanishes.

The same coefficient matching extends order by order in $k_L$. Denoting the corresponding terms at fixed $(\mathsf{c}_1,\mathsf{c}_2)$ by $\mathcal T_{N,\mathrm{cut};\mathsf{c}_1\mathsf{c}_2}$ and $\mathcal T_{2,\mathrm{cut};\mathsf{c}_1\mathsf{c}_2}^{(r,s)}$, we obtain
\begin{align}
  \mathcal T_{N,\mathrm{cut};\mathsf{c}_1\mathsf{c}_2} =\sum_{r,s=0}^{\infty}g^{(r,s)}_{\mathsf{c}_1\mathsf{c}_2} \mathcal T_{2,\mathrm{cut};\mathsf{c}_1\mathsf{c}_2}^{(r,s)}\,. \label{matching}
\end{align}
Every auxiliary bubble $\mathcal T_2^{(r,s)}$ satisfies case (i) of the exclusion rule, so its CC signal vanishes. The matching relation in \eqref{matching} then implies that the $N$-gon CC signal vanishes to all soft orders.

\section{Triangle with a chemical potential}
\label{sec:triangle}

For products and ratios of Gamma functions in this section, we use the shorthand notations
\begin{align}
  \Gamma\left[a_1,\ldots,a_n\right]&\equiv\prod_{i=1}^{n}\Gamma(a_i)\,,\qquad \Gamma\left[\begin{matrix}a_1,\ldots,a_n\\b_1,\ldots,b_m\end{matrix}\right]\equiv\frac{\prod_{i=1}^{n}\Gamma(a_i)}{\prod_{j=1}^{m}\Gamma(b_j)}\,.
\end{align}

For $\nu>0$, the normalised BD solution of \eqref{Whittaker} is
\begin{align}
  v(k,\tau) =\frac{e^{\pi\kappa/2}}{\sqrt{2k}} \mathrm{W}_{-\ii\kappa,\,\ii\nu}(2\ii k\tau)\,, \label{v}
\end{align}
up to an irrelevant constant phase. The physical mode is then $u(k,\tau)=v(k,\tau)/a(\tau)$, with $a(\tau)=-1/\tau$. In particular, its leading late-time expansion can be written as
\begin{align}
  \lim_{-k\tau\to 0} u(k,\tau)= e^{-\ii\pi/4}\frac{e^{\pi\kappa/2}}{\sqrt{k^3}} \left[ C_+(\kappa,\nu)(-k\tau)^{3/2+\ii\nu} +C_-(\kappa,\nu)(-k\tau)^{3/2-\ii\nu} \right]\,, \label{lateTime}
\end{align}
where the coefficients \(C_\pm(\kappa,\nu)\) are given by
\begin{align}
  C_+(\kappa,\nu)=\frac{2^{\ii\nu}e^{\pi\nu/2}\,\Gamma(-2\ii\nu)}{\Gamma{(\frac12-\ii\nu+\ii\kappa})}\,,\qquad C_-(\kappa,\nu)=C_+(\kappa,-\nu)\,.
\end{align}

For $\kappa=0$, $u$ reduces to the Hankel mode in \eqref{Hankel} up to a constant phase. The SK propagators are given by \eqref{Dmp}-\eqref{Dmm} with this mode, and their dependence on $(\kappa,\nu)$ is suppressed. For the conformally coupled external scalar, the boundary propagator is
\begin{align}
  K_{\mathsf a}(k,\tau) =\frac{\tau_f\tau}{2k}e^{\ii\mathsf a k\tau}\,, \qquad \mathsf a=\pm1\,,
\end{align}
to leading order as $\tau_f\to0$.

Three insertions of the interaction in \eqref{interaction} give
\begin{align}
  \left\langle \phi_{\bm k_1}\phi_{\bm k_2}\phi_{\bm k_3} \right\rangle'_{\tau_f} &= \frac{\lambda^3\tau_f^3}{8k_1k_2k_3} \sum_{\mathsf a_1,\mathsf a_2,\mathsf a_3=\pm1} \left[\prod_{v=1}^{3}\int_{-\infty}^{\tau_f} \frac{-\ii\mathsf a_v\dd\tau_v}{\tau_v^3} e^{\ii\mathsf a_vk_v\tau_v}\right] \nonumber\\
  &\times\int\frac{\dd^3\bm q}{(2\pi)^3}\, D_{\mathsf a_1\mathsf a_3}(q;\tau_1,\tau_3) D_{\mathsf a_3\mathsf a_2}(|\bm q+\bm k_3|;\tau_3,\tau_2) D_{\mathsf a_1\mathsf a_2}(|\bm q-\bm k_1|;\tau_1,\tau_2)\,, \label{triangle}
\end{align}
where $\bm k_1+\bm k_2+\bm k_3=0$ and the momentum convention agrees with Fig.~\ref{fig:selection-rule}(d).

It is straightforward to show that the directional cutting rule of Sec.~\ref{sec:cutting} also extracts the leading CC signal for Whittaker modes. In the squeezed limit $k_3=k_L\ll k_S=k_1+k_2$, we repeat the steps leading to \eqref{factorisation} for $N=3$, using the Whittaker mode and its late-time expansion in \eqref{v} and \eqref{lateTime}. Retaining the leading terms $m=n=0$ with equal sign branches, setting $k_1=k_2=k_S/2$ at this order, and using the normalisation in \eqref{bispectrum}, we obtain
\begin{align}
  &\lim_{k_L/k_S\to 0} B_{\mathrm{CC}} = \frac{e^{2\pi\kappa}k_S^2}{4k_L^2} \sum_{\mathsf a_1,\mathsf a_2=\pm1} (-\mathsf a_1\mathsf a_2) \int_{-\infty}^{0} \dd\tau_1\dd\tau_2\,(-\tau_1)^{-3/2+\ii\nu}(-\tau_2)^{-3/2+\ii\nu} D_{\mathsf a_1\mathsf a_2}(k_S/2;\tau_1,\tau_2) e^{\ii k_S(\mathsf a_1\tau_1+\mathsf a_2\tau_2)/2} \nonumber\\
  &\times \bigg[ C_+^2(\kappa,\nu) \int_{-\infty}^{0} \frac{-\dd\tau_3}{\tau_3}\, e^{\ii k_L\tau_3} \int\frac{\dd^3\bm q}{(2\pi)^3}\, q^{-1/2+\ii\nu} |\bm q+\bm k_L|^{-1/2+\ii\nu} \mathrm{W}_{\ii\kappa,\ii\nu}(-2\ii q\tau_3) \mathrm{W}_{\ii\kappa,\ii\nu} (-2\ii|\bm q+\bm k_L|\tau_3)\nonumber\\
  &+ C_+^{*2}(\kappa,-\nu) \int_{-\infty}^{0} \frac{-\dd\tau_3}{\tau_3}\, e^{-\ii k_L\tau_3} \int\frac{\dd^3\bm q}{(2\pi)^3}\, q^{-1/2+\ii\nu} |\bm q+\bm k_L|^{-1/2+\ii\nu} \mathrm{W}_{-\ii\kappa,\ii\nu}(2\ii q\tau_3) \mathrm{W}_{-\ii\kappa,\ii\nu} (2\ii|\bm q+\bm k_L|\tau_3) \bigg] +\text{c.c.} \label{BCC}
\end{align}

The hard time factor in \eqref{BCC} is a tree-level Whittaker two-point seed integral. Following \cite{Qin:2023ejc}, we define its dimensionless form and quote the corresponding closed-form result:
\begin{align}
  \mathcal I(\kappa,\nu)&\equiv k_S^{2+2\ii\nu}\sum_{\mathsf a_1,\mathsf a_2=\pm1} (-\mathsf a_1\mathsf a_2) \int_{-\infty}^{0} \dd\tau_1\dd\tau_2\, (-\tau_1)^{-3/2+\ii\nu} (-\tau_2)^{-3/2+\ii\nu} D_{\mathsf a_1\mathsf a_2}(k_S/2;\tau_1,\tau_2) e^{\ii k_S(\mathsf a_1\tau_1+\mathsf a_2\tau_2)/2}\nonumber\\
  &= \Gamma(1+2\ii\nu) \left( \frac{\ii e^{\pi\nu}} {\frac12+\ii\nu-\ii\kappa} - \frac{\ii e^{-\pi\nu}} {\frac12+\ii\nu+\ii\kappa}\right) + \frac{e^{\pi\kappa}}{(\frac12+\ii\nu-\ii\kappa)(\frac12+\ii\nu+\ii\kappa)}\Gamma\left[\begin{matrix} 2+2\ii\nu,2+2\ii\nu\\ \frac32+\ii\nu-\ii\kappa,\frac32+\ii\nu+\ii\kappa \end{matrix}\right] \,. \label{I}
\end{align}

It remains to evaluate the two combined time and loop integrals in the square brackets of \eqref{BCC}. For later convenience, we define
\begin{align}
  \mathcal J_\pm(\kappa,\nu) &\equiv k_L^{-2-2\ii\nu} \int_{-\infty}^{0}\frac{-\dd\tau_3}{\tau_3}\, e^{\pm\ii k_L\tau_3} \nonumber\\
  &\times\int\frac{\dd^3\bm q}{(2\pi)^3}\, q^{-1/2+\ii\nu} |\bm q+\bm k_L|^{-1/2+\ii\nu} \mathrm{W}_{\pm\ii\kappa,\ii\nu}(\mp2\ii q\tau_3) \mathrm{W}_{\pm\ii\kappa,\ii\nu} (\mp2\ii|\bm q+\bm k_L|\tau_3)\,. \label{J}
\end{align}
To evaluate this integral, we use the MB representation of the Whittaker functions:
\begin{align}
  \mathrm{W}_{\pm\ii\kappa,\ii\nu}(\mp 2\ii k\tau) &= \int_{-\ii\infty}^{+\ii\infty}\frac{\dd s}{2\pi\ii}\,e^{\pm\ii\pi/4\mp\ii\pi s/2}(-2k\tau)^{1/2-s}\mathcal K_\pm(s)\,,\\
  \mathcal K_\pm(s)&\equiv \Gamma\left[s-\ii\nu,s+\ii\nu\right] \pregFq{2}{1}{s-\ii\nu,s+\ii\nu}{s\mp\ii\kappa+\frac12}{\frac12}\,,
\end{align}
where
\begin{align}
  \pregFq{2}{1}{a,b}{c}{z} \equiv \frac{1}{\Gamma(c)}\pFq{2}{1}{a,b}{c}{z}
\end{align}
is the regularised hypergeometric function that is analytic for all indices $a,b,c$ at $z=1/2$. Substituting the MB representation into \eqref{J}, we can perform both the time integral and the loop momentum integral analytically. Writing \(S\equiv s_1+s_2\), the time integral gives
\begin{align}
  \int_{-\infty}^0\dd\tau_3\,(-\tau_3)^{-S}e^{\pm\ii k_L\tau_3} =\mp\ii e^{\pm\ii\pi S/2} \Gamma(1-S)k_L^{S-1}\,,
\end{align}
while the loop momentum integral is
\begin{align}
  \int\frac{\dd^3\bm q}{(2\pi)^3}\, q^{\ii\nu-s_1} |\bm q+\bm k_L|^{\ii\nu-s_2} = \frac{k_L^{3+2\ii\nu-S}}{(4\pi)^{3/2}} \Gamma\left[\begin{matrix} \frac{S-3-2\ii\nu}{2},\frac{3+\ii\nu-s_1}{2},\frac{3+\ii\nu-s_2}{2}\\ \frac{s_1-\ii\nu}{2},\frac{s_2-\ii\nu}{2},3+\ii\nu-\frac{S}{2} \end{matrix}\right]\,.
\end{align}
Combining the two results, the dependence on \(k_L\) factorises completely, leaving a dimensionless double MB integral
\begin{align}
  \mathcal J_\pm(\kappa,\nu) = \frac{1}{(4\pi)^{3/2}} \int_{-\ii\infty}^{+\ii\infty}\frac{\dd s_1\dd s_2}{(2\pi\ii)^2}\,2^{1-S}\mathcal K_\pm(s_1)\mathcal K_\pm(s_2) \Gamma\left[\begin{matrix} 1-S,\frac{S-3-2\ii\nu}{2},\frac{3+\ii\nu-s_1}{2},\frac{3+\ii\nu-s_2}{2}\\ \frac{s_1-\ii\nu}{2},\frac{s_2-\ii\nu}{2},3+\ii\nu-\frac{S}{2} \end{matrix}\right]\,.
\end{align}

The remaining double MB integral can be evaluated analytically. Using the result in our companion work, we obtain the following representation in terms of Meijer-\(\mathrm{G}\) and hypergeometric functions:
\begin{align}
  \mathcal J_\pm(\kappa,\nu) &= \kappa \Gamma\left[-3-2\ii\nu,\frac12+\ii\nu\pm\ii\kappa\right] \Bigg\{ \frac{ \pm\ii 2^{-2-2\ii\nu\mp2\ii\kappa} (1\pm2\ii\kappa) }{ \pi^{3/2} \left(\frac12+\ii\nu\pm\ii\kappa\right) } \Gamma\left[\begin{matrix} 1+2\ii\nu\\2+\ii\nu\pm\ii\kappa,1\mp2\ii\kappa \end{matrix}\right] \nonumber\\
  &\quad\times \Bigg[ -4\kappa^2 \pFq{2}{1}{1,-2-2\ii\nu\mp2\ii\kappa}{-1\mp2\ii\kappa}{-1} +2\left(\frac12+\ii\nu\right)^2+2\kappa^2 - \frac{ \left(\frac12+\ii\nu\mp\ii\kappa\right)^2 (1+2\ii\nu)}{-1\mp2\ii\kappa}\Bigg] \nonumber\\
  &+ \frac{2^{-1+2\ii\nu}}{\pi^3\Gamma\!\left(\frac12-\ii\nu\mp\ii\kappa\right)} \biggl\{ \left[ \pm2\ii\left(\frac12+\ii\nu\pm\ii\kappa\right)(-3-2\ii\nu)-8\kappa\right] \nonumber\\
  &\qquad\times\MeijerG{4}{3}{5}{5}{1,2,\frac32;\frac52+\ii\nu+\ii\kappa,\frac52+\ii\nu-\ii\kappa}{\frac32+\ii\nu,2+\ii\nu,2,-\frac12-\ii\nu\mp\ii\kappa;\frac52+\ii\nu\pm\ii\kappa}{1} \nonumber\\
  &\qquad + 4\kappa\MeijerG{4}{3}{5}{5}{1,2,\frac32;\frac52+\ii\nu+\ii\kappa,\frac52+\ii\nu-\ii\kappa}{\frac32+\ii\nu,2+\ii\nu,3,-\frac12-\ii\nu\mp\ii\kappa;\frac52+\ii\nu\pm\ii\kappa}{1} \biggr\} \Bigg\}\,. \label{Jexact}
\end{align}
For numerical evaluation, a simpler representation follows by reducing the two Mellin integrals to a single ordinary integral:
\begin{align}
  \mathcal J_\pm(\kappa,\nu) &= \frac{-\ii}{2^{6+2\ii\nu}\pi^3} \Gamma\left[\begin{matrix} 2+2\ii\nu,2+2\ii\nu,-3-2\ii\nu\\ \frac52+\ii\nu\mp\ii\kappa,\frac52+\ii\nu\mp\ii\kappa \end{matrix}\right] \int_0^\infty\frac{\dd z}{z}\, \left[ \left(1+\frac{4\ii}{z}\right)^{3+2\ii\nu}- \left(1-\frac{4\ii}{z}\right)^{3+2\ii\nu} \right] \nonumber\\
  &\times \left( \pFq{2}{1}{2+2\ii\nu,2}{\frac52+\ii\nu\mp\ii\kappa}{\frac12+\frac{2\ii}{z}} - \pFq{2}{1}{2+2\ii\nu,2}{\frac52+\ii\nu\mp\ii\kappa}{\frac12-\frac{2\ii}{z}} \right)^2\,.
\end{align}

Combining \eqref{BCC}, \eqref{I}, and \eqref{J}, we finally obtain the leading CC signal
\begin{align}
  \lim_{k_L/k_S\to 0} B_{\mathrm{CC}} &= \mathcal A(\kappa,\nu)\left(\frac{k_L}{k_S}\right)^{2\ii\nu} + \text{c.c.}\,,\\
  \mathcal A(\kappa,\nu) &\equiv \frac{e^{2\pi\kappa}}{4}\mathcal I(\kappa,\nu)\left[C_+^2(\kappa,\nu)\mathcal J_+(\kappa,\nu) + C_+^{*2}(\kappa,-\nu)\mathcal J_-(\kappa,\nu)\right]\,.
\end{align}
The characteristic nonanalytic scaling $\left(k_L/k_S\right)^{2\ii\nu}$ exhibits the expected CC oscillation with frequency \(2\nu\). In particular, \eqref{Jexact} gives
\begin{align}
  \mathcal A(0,\nu)=0\,.
\end{align}
Thus the candidate massive late-time scaling remains present in the individual heavy field propagators, while its nonanalytic coefficient vanishes when dS covariance is restored, providing an explicit one-loop realisation of the exclusion rule discussed in the main text.
\end{document}